\documentclass[twocolumn]{aastex63}

\usepackage{booktabs}

\turnoffedit

\accepted{September 22, 2022}
\shorttitle{Cepheus Far North}
\shortauthors{Kerr et al.}

\begin{document}

\title{SPYGLASS. II. The Multi-Generational and Multi-Origin Star Formation History of Cepheus Far North}

\correspondingauthor{Ronan Kerr}
\email{rmpkerr@utexas.edu}

\author[0000-0002-6549-9792]{Ronan Kerr}
\affiliation{Department of Astronomy, University of Texas at Austin\\
2515 Speedway, Stop C1400\\
Austin, Texas, USA 78712-1205\\}

\author[0000-0001-9811-568X]{Adam L. Kraus}
\affiliation{Department of Astronomy, University of Texas at Austin\\
2515 Speedway, Stop C1400\\
Austin, Texas, USA 78712-1205\\}

\author[0000-0002-5648-3107]{Simon J. Murphy}
\affiliation{Centre for Astrophysics, University of Southern Queensland\\
Toowoomba, QLD 4350 Australia.\\}

\author[0000-0001-9626-0613]{Daniel M. Krolikowski}
\affiliation{Department of Astronomy, University of Texas at Austin\\
2515 Speedway, Stop C1400\\
Austin, Texas, USA 78712-1205\\}

\author[0000-0003-1252-9916]{Stella S. R. Offner}
\affiliation{Department of Astronomy, University of Texas at Austin\\
2515 Speedway, Stop C1400\\
Austin, Texas, USA 78712-1205\\}

\author[0000-0003-2053-0749]{Benjamin M. Tofflemire}
\affiliation{Department of Astronomy, University of Texas at Austin\\
2515 Speedway, Stop C1400\\
Austin, Texas, USA 78712-1205\\}

\author[0000-0001-9982-1332]{Aaron C. Rizzuto}




\begin{abstract}

Young stellar populations provide a record of past star formation, and by establishing their members’ dynamics and ages, it is possible to reconstruct the full history of star formation events. Gaia has greatly expanded the number of accessible stellar populations, with one of the most notable recently-discovered associations being Cepheus Far North (CFN), a population containing hundreds of members spanning over 100 pc. With its proximity (d~$\la$~200~pc), apparent substructure, and relatively small population, CFN represents a manageable population to study in depth, with enough evidence of internal complexity to produce a compelling star formation story. Using Gaia astrometry and photometry combined with additional spectroscopic observations, we identify over 500 candidate CFN members spread across 7 subgroups. Combining ages from isochrones, asteroseismology, dynamics, and lithium depletion, we produce well-constrained ages for all seven subgroups, revealing a largely continuous 10 Myr star formation history in the association. By tracing back the present-day populations to the time of their formation, we identify two spatially and dynamically distinct nodes in which stars form, one associated with $\beta$ Cephei which shows mostly co-spatial formation, and one associated with EE Draconis with a more dispersed star formation history. This detailed view of star formation demonstrates the complexity of the star formation process, even in the smallest of regions. 
\end{abstract}



\section{Introduction} \label{sec:intro}

Most local star formation leaves behind young associations, unbound stellar groupings that inherit their dynamics from the dense clouds that they emerged from \citep[e.g.,][]{Krumholz19, Krause20}. These associations can act as a stellar fossil record, holding an imprint of the entire star formation event, spanning timescales that cannot be investigated through studies of gas dynamics in sites of active star formation \citep{Briceno07}. Through the acquisition of only the space-velocity configurations and ages of members, it is possible to use stellar populations to provide a detailed view of an association's star formation history \citep[e.g., see][]{Kounkel18,MiretRoig20}, revealing not only which populations are most closely related to one another but also how subgroups interacted throughout the formation process. 

Numerous properties of the parent cloud can be uncovered by using traceback to reconstruct star formation. Current and future star-forming clouds host a variety of different structures, spanning from long isolated filaments to the more centralized and high-density star formation in places like $\rho$ Oph and Orion \citep[e.g.,][]{Zucker15, Kirk17, Kerr19}. The use of traceback on stellar populations allows for the reconstruction of these structures, which in turn provides critical priors on models for the assembly of stellar populations out of a parent cloud \citep[e.g.,][]{Grudic21, Guszejnov22}. Beyond revealing whether the progenitor cloud is spherical or filamentary in form, the distribution of subgroups at the times of formation can also reveal whether star formation was scattered or localized in a small number of collection hubs. The presence of a small number of common sites of star formation in an otherwise extended  structure would be consistent with the filamentary accretion model of star formation, in which gas flows along filaments to collection hubs, producing a small number of more spherical clusters out of a parent filament \citep[e.g.,][]{Kirk13, Krause20}. The distribution of ages within structures reveals whether subsequent generations were continuous with one another or defined by bursts separated by pauses. Stellar subgroups with large age separations have been proposed as indicative of star formation disruption by stellar feedback, therefore potentially providing an important window into these processes \citep[e.g.,][]{Beccari17,Kerr21}. The overall range of ages also provides important insight into the timescale of star formation, which is an important observable in many current simulations \citep[e.g.,][]{Grudic21}. 

There is a long history of detailed studies investigating individual young associations, with previous studies providing results that include age estimations through a variety of methods and traceback which help to guide the reconstruction of star formation histories \citep[e.g.,][]{Pecaut16, Krause18, CantatGaudin19, Krolikowski21}. However, until very recently, our astrometric coverage of these populations lacked completeness. This incomplete dynamical record prevented measuring most stellar 3D or even 2D motions, which has the effect of reducing the number of stars available not just for performing dynamical studies but also for identifying more tenuous subgroups. As a result, association-level studies have until recently been significantly limited in depth, and have only covered populations that are both large and well-established, such as the Sco-Cen association, Taurus, and Orion \citep[e.g.,][]{Brown94, deZeeuw99, Pecaut16, Kraus17}. The recent results of the Gaia spacecraft have provided proper motions and distances for nearly 2 billion stars \citep{GaiaDR218, EDR3Astro_Lindegren21}, providing the expansion to our dynamical coverage of nearby stellar populations necessary to not just reveal new associations, but also to perform detailed dynamical studies. Since Gaia data releases began, multiple new catalogs of stellar populations have been released, including populations of all ages, from old and bound open clusters to the young associations that interest us \citep[e.g.,][]{Sim19, Kounkel19}.

One of the most detailed studies of young populations ($\tau<50$ Myr) was produced by \citet{Kerr21} (hereafter SPYGLASS-I), which used a Bayesian framework to identify young stars and exclude older populations, allowing detailed clustering which can identify young associations. A total of 27 top-level associations were identified through SPYGLASS-I, many of which were little-known or completely absent from the literature. This rich collection of little-studied associations therefore produces a valuable sample of young stars with extensive Gaia dynamical coverage, revealing star formation histories in environments unlike any prior work. 

Of all the young associations that have begun to emerge through SPYGLASS-I and similar searches, the Cepheus Far North association (CFN) is one of the largest and most accessible. Until very recently, this association had been considered as little more than a limited set of young stars in the foreground of the better-studied and more distant star-forming environments in Cepheus Flare \citep[e.g., ][]{Tachihara05, Klutsch11, Oh17, Frasca18}. It was known as the ``Cepheus Association'', a name that SPYGLASS-I updated to Cepheus Far North (CFN) to avoid confusion with other young associations in and around Cepheus. The first paper to perform a detailed and targeted analysis of CFN was \citet{Klutsch20}, which identified only 32 candidate members limited to the dense region centered at a distance of 157 pc. The extent of known populations in the vicinity of CFN's core was expanded by \citet{Szilagyi21}, who identified new candidate members using literature young star lists including \citet{Zari18}. This publication also revealed a second population in the region, which they referred to as the HD 190833 association. Between these two components, \citet{Szilagyi21} more than tripled the population identified in \citet{Klutsch20}, identifying 37 new candidate members in the CFN core and another 46 associated with HD 190833.

SPYGLASS-I also considerably expanded the known membership of CFN, identifying the population as a large stellar overdensity containing 219 photometrically young stars, distributed over an area more than 100 pc across \edit1{and divided into two subgroups. The main group identified, CFN-1, merges CFN's core with the HD 190833 association, and contains a majority of the population of the association.} Despite the extensive size and irregular shape of CFN-1, the HDBSCAN clustering employed in SPYGLASS-I viewed the entire region from the CFN Core to HD 190833 as a single, contiguous group with no discernible lower-level substructure like that identified in \citet{Szilagyi21}. The other subgroup identified by SPYGLASS-I, CFN-2, is a more distant population containing $\beta$ Cephei that has not been recognized in any other publications. CFN's irregular shape, emerging substructure, and large size may nonetheless suggest the presence of multiple spatially distinct episodes of star formation that may emerge in a more detailed kinematic analysis.
The recent discovery of CFN \edit1{therefore} provides a unique opportunity to gain insight into the dynamics of smaller associations far from the influence of larger and better-studied environments like Sco-Cen, Taurus, and Orion. 


In this paper, we perform the first detailed dynamical study of CFN, while greatly expanding our known populations in the region. In Section \ref{sec:candidateselec}, we outline our selection of candidate members. In Section \ref{sec:datacomp}, we outline observations and external data collection undertaken to provide radial velocities and youth indicators to supplement Gaia astrometric data. Using our combined data set, we finalize details of the membership in Section \ref{sec:membership}. Our analysis of the association's properties and history is found in Section \ref{sec:results}, which includes a new and detailed view of the substructure and ages of stars in CFN, before tracing stars back through the entire star formation history in Sco-Cen. In Section \ref{sec:discussion}, we discuss the implications of the star formation patterns in CFN, before concluding in Section \ref{sec:conclusion}. 

\section{Candidate Selection} \label{sec:candidateselec}

Our selection of young stars in CFN is based on the census of young stars and associations from SPYGLASS-I, which revealed over 3$\times$10$^4$ photometrically young stars within 333 pc of the solar system. The HDBSCAN clustering algorithm identified 219 young stars as part of the association, making it one of the largest groups in that sample outside of the well-known large associations like Sco-Cen, Orion, Taurus, and Vela. However, the sample of young stars in SPYGLASS-I was based primarily on stars with robust Gaia astrometry that reside high on the pre-main sequence. Therefore, many genuine members of the association will have been missed, mainly early-type stars that are already on the main sequence. With a SPYGLASS-I age of approximately 24 Myr, stellar recovery rates of under 50\% are expected in CFN, so for a complete sample it is important that kinematic and spatial neighbors to the young sample also be vetted for potential membership. We can also improve the sample of CFN members by updating that analysis to the \edit1{Gaia EDR3 data set. EDR3's photometry and astrometry is largely the same as what was recently published as part of the full DR3 release, which provided some minor corrections to $m_G$ \citep{GaiaDR3_22}. However, in the relevant set for this analysis we found changes of $\Delta m_G \la 0.025$ affecting approximately 10\% of the data set, making the effects of a further update to DR3 minimal. These EDR3 measurements nonetheless provide} a significant improvement to the photometric and astrometric quality compared to the Gaia DR2 data set used in SPYGLASS-I. 

To broaden and enrich our coverage of CFN, we first revised our photometrically young sample by re-applying the SPYGLASS-I stellar population identification methods to the new Gaia EDR3 data set, including the script for Bayesian photometric identification of young stars and the HDBSCAN clustering routine. We kept the same quality restrictions used in that publication to exclude less reliable Gaia astrometric or photometric solutions. These restrictions included a photometric quality cut based on the BP/RP flux excess factor, which flags errant flux in the BP and RP color bands versus the main Gaia G band, an astrometric cut based on the unit weight error, which can be viewed as an astrometric goodness of fit parameter, and a requirement that $\pi$/$\sigma_\pi > 10$, which ensures reliable distances. SPYGLASS-I also used a cut on the number of visibility periods used in the astrometric solution, however with the improved coverage in EDR3 we found that no stars in our sample failed this cut, so it was removed. The exact restrictions are provided in SPYGLASS-I, which are a modified form of those proposed in \citet{Arenou18}. When the Bayesian photometric youth script and HDBSCAN clustering was applied, we found that our EDR3-updated data set simultaneously expanded the sample of likely young members to 222 while also refining the group's extent, excluding a few outlying SPYGLASS-I members that were found to be too far from other members in this revised population. 

To identify spatial and kinematic neighbors that may represent additional members, we applied the method described in SPYGLASS-I for identifying neighbors. We selected all stars with a distance to the 10th nearest photometrically young candidate member (d$_{10}$) consistent with the range possessed by those photometrically young candidates, and assigned each star a ``clustering proximity'' parameter, $D$ (previously referred to as ``strength'' in SPYGLASS-I), between zero and one, where $D=0$ corresponds to the largest d$_{10}$ for a photometrically young candidate, and $D=1$ corresponds to the smallest. $D$ behaves similarly to the cube root of density, as it has an inverse relation to the length scale of a cube that contains a fixed number of stars. To prioritize completeness over purity, we used looser quality restrictions on the stars assessed as possible candidate members compared to the initial young sample used to define the extent of the group during clustering. As such, only stars that lacked a 5-parameter astrometric solution or a G magnitude were removed from this sample. This ensured that all stars have the two-dimensional velocity vector necessary to assess common motions, and the magnitudes necessary to assess the feasibility of RV follow-up. Through robust spectroscopic follow-up we can determine whether radial velocities are consistent with membership and search for youth indicators like Li, H-alpha emission, and fast rotation, producing new membership criteria that can easily compensate for the looser initial selection criteria, especially for stars in a region of the color-magnitude diagram (CMD) without strong youth diagnostics. We preserved the metrics used to check quality in SPYGLASS-I and for the young sample in the form of astrometric and photometric quality flags, which are available in Table \ref{tab:members} for if a high-quality subset of candidate members is desired. The astrometric flag provides the star's boolean solution to our unit weight error-based cut, while the photometric flag provides the boolean solution to the BP/RP flux excess factor-based cut. 

The resulting expanded population of candidate members contained 2484 objects. However, upon investigating the sample as a function of $D$, we found that only about 5\% of stars below the pre-main sequence turn-on with $D<0.05$ had photometry consistent with likely membership (as defined in Section \ref{sec:photselec}), a fraction low enough that field binaries may begin to dominate that low-$D$parameter space \citep[e.g.,][]{Sullivan21}. The presence of unrelated binaries in the crowded field around CFN in turn likely caused the original HDBSCAN-defined extent of CFN to be larger than it should be, as the field binaries can merge with the pre-main sequence and inflate the apparent occurrence of young stars on the group's edge. Requiring that $D>0.05$ removes objects in the field-dominated outer reaches of CFN, restricting the list of 901 candidates. While there are likely some CFN members beyond this limit, they are unlikely to merit the observing time required to distinguish members from field interlopers. The extent of members that may have been missed is discussed in Section \ref{sec:completeness}.

\section{Observations and Literature Data} \label{sec:datacomp}

\subsection{Literature Radial Velocities}

Radial velocities are critical for both confident determinations of membership and high-quality kinematic studies in 3 dimensions. \edit1{Gaia DR3 provides very broad RV coverage of CFN candidates, and we therefore added these values to the Gaia EDR3-based data set used for kinematics and photometry.} While Gaia DR3 typically reported radial velocities for stars with G $\lesssim$ \edit1{14}, covering \edit1{287} out of the total CFN candidate list of 901, only \edit1{67} of those have a sub-km s$^{-1}$ uncertainty. To improve the completeness of our radial velocity sample, we collected additional sources from Simbad and Vizier, keeping the lowest-uncertainty value from literature. \edit1{We also excluded or removed RVs with uncertainties greater than half the $\sim$7 km s$^{-1}$ maximum radius of CFN in transverse velocity space (see Sec \ref{sec:rv_selec}), ensuring that all stars can have their membership assessed at a 2$\sigma$ level. The resulting search} provided an additional \edit1{8} measurements from external sources \edit1{\citep{Bobylev06, Gontcharov06, Kharchenko07, Frasca18}}, replacing a lower-quality Gaia measurement in \edit1{two cases. This brought the }total number of literature radial velocity measurements in CFN to \edit1{160, with 152 Gaia DR3 measurements still being used after the $\sigma_{v_r}$ cut and the introduction of external observations}. Radial velocity and other stellar properties are compiled in Tables \ref{tab:spectres} and \ref{tab:members}, with Table \ref{tab:spectres} covering RV and spectral line data for the subset of stars where it is available, and Table \ref{tab:members} covering the complete sample of credible CFN members, alongside some properties and flags that are available for all members. 

Two literature radial velocity observations are ambiguously attributed to pairs resolved in Gaia, both of which form near-equal brightness binaries in which both components would be expected to contribute similarly to the RV measurements. In these cases, the observation is attributed to both objects, and measurements are flagged accordingly in Table \ref{tab:members}. 
All literature radial velocities that are used in our final data set are included in Table \ref{tab:spectres}, together with their source, although only \edit1{73 of the 160 literature measurements} are not overwritten by higher-quality radial velocity measurements gathered through spectroscopy that are presented in Section \ref{sec:obs}. 

\subsection{New Spectroscopic Observations} \label{sec:obs}

While most literature RV measurements are sufficient to assess a star's membership in CFN, approximately \edit1{32\%} of candidate member stars brighter than magnitude \edit1{14} lack \edit1{a credible} radial velocity measurement, and the measurements that do exist are often not sufficiently precise for \edit1{traceback. Stars with $\sigma_{v_R} < 1$ km s$^{-1}$ are generally necessary for accurate traceback in this instance, as a 1 km s$^{-1}$ deviation diverges from the true value at a rate of $\sim$1 pc Myr$^{-1}$. Assuming SPYGLASS-I's age estimate of $\tau\sim24$ Myr, a 1 km s$^{-1}$ uncertainty would produce deviations slightly smaller than the 30-40 pc scales of the visible CFN substructure during the time since formation, making this uncertainty an approximate upper limit for enabling convergence upon traceback.} Gaia radial velocities in particular often show inconsistencies with those of our own independent observations at the km s$^{-1}$ level, even for objects with sub-km s$^{-1}$ Gaia uncertainties, strengthening our motivation for the widespread coverage of the association with the high-precision, accurate, and self-consistent spectroscopic measurements that can facilitate kinematic studies. New spectral observations can be useful even for objects that already have reliable radial velocity measurements, as youth indicators like hydrogen emission and lithium absorption can be used to independently verify a star's young age.

For the purpose of acquiring precise dynamical measurements in young stars, the best radial velocities generally come from later-type stars that have less severe line broadening as a result of their slower rotation \citep[e.g.,][]{Rebull20, Bouma22}. However, for the purposes of establishing membership, RV coverage of the bright candidate members is much more important. Only stars with an absolute magnitude $M_G>6.5$ (corresponding to an apparent magnitude $m_G\gtrsim12$ at the distance of CFN, or M $\la 0.8$ M$_{\odot}$) can be distinguished between the pre-main sequence of the young association and the older sequence of the field stars. We therefore employ observations from two different spectrographs to obtain improved radial velocity measurements which both complete the radial velocity coverage of stars no longer on the pre-main sequence and provide access to spectroscopic youth indicators. 

Most observations used the TS23 configuration of the Robert G. Tull Coud\'e spectrograph at the McDonald Observatory's 2.7m Harlan J. Smith Telescope (HJST), which provides high-resolution spectra with $R=60000$ for a spectral range between 3400 and 10900 \AA ~\citep{Tull95}. Our observations spanned two programs. The first program targeted stars with $m_G<12$ and a clustering proximity of $D>0.25$, while the second observed a broader selection ($D>0.05$) of later-type stars on the pre-main sequence with magnitudes $12<m_G<14$, a section of the CMD that reliably produces high-quality RV measurements alongside clear spectral youth indicators. The two programs therefore served the complementary purposes of first building our coverage of the association above the pre-main sequence turn-on for membership purposes, and second providing a deep sample of high-quality radial velocities for kinematic studies. In total, 97 spectra were taken over the course of 14 nights covering 94 targets, with a few duplicate observations in cases where an initial observation was made under poor conditions. 

Our exposure times ranged from 5 to 30 minutes depending on the stellar magnitude, aiming for S/N ratios of at least 30 around the Li 6708 \AA~ line, which enables both radial velocity measurements with sub-km s$^{-1}$ and the robust detection and measurement of Li equivalent widths. However, the signal to noise ratio was allowed to drop closer to 10 in cases where a target was too dim to hit our signal objective in 30 minutes of exposure, providing spectra in which sub-km s$^{-1}$ radial velocities are still readily attainable, but individual lines like Li can be more tenuous. H$\alpha$ was on the detector for only 26\% of our observations, as we often excluded it in favor of centering the blaze peak and maximizing overall signal. For that reason, we used H$\beta$ instead of H$\alpha$ as our tracer of Hydrogen emission. Spectra were reduced using a custom python implementation of standard reduction procedures. After bias subtraction, flat-field correction, and cosmic ray rejection, we extracted 1D spectra from the 2D echellograms using optimal extraction. We derived wavelength solutions using a series of ThAr lamp comparison observations taken throughout each observing night.

The remaining observations used the NRES spectrographs at the 1m nodes of the Las Cumbres Observatory network (LCO), which provide high-resolution spectra with $R=53000$ for a spectral range between 3800 and 8600 \AA. These observations primarily targeted brighter targets at lower $D$ values not reached during the first allocation at HJST. Like at HJST, we aimed for a S/N ratio around 30 for the Li spectral window for easy observation of the Li 6708 \AA~ line, allowing the S/N ratio to drop closer to 10 for stars at very low end of our magnitude range, which for NRES was limited to $m_G<13$. This meant exposure times of 8 minutes for the brightest targets, and 30 minutes for the faintest. Due to the near complete spectral coverage of NRES within its advertised wavelength range, H$\alpha$ was covered in all observations at the LCO, as was H$\beta$. A total of 19 targets were observed using NRES, spread between the Wise Observatory (TLV) and McDonald Observatory (ELP) nodes. This LCO set completed our RV coverage of bright objects (G$<$6.5) in CFN with $D>0.05$, excluding only a handful of objects that either fail the Gaia astrometric quality flags which make them ineligible for RV-based membership assessment, or have evidence for the presence of an unresolved companion which is likely to undermine the reliability of the Gaia astrometric solution (see Section \ref{sec:sss}). NRES data products are automatically reduced using the BANZAI data reduction pipeline \citep{McCullyBANZAI18}, which extracts each order and wavelength-calibrates the solution. As such, these spectra are ready-to-use for RV and equivalent width measurements upon download from the LCO archive. 

\begin{figure}
\centering
\includegraphics[width=8.2cm]{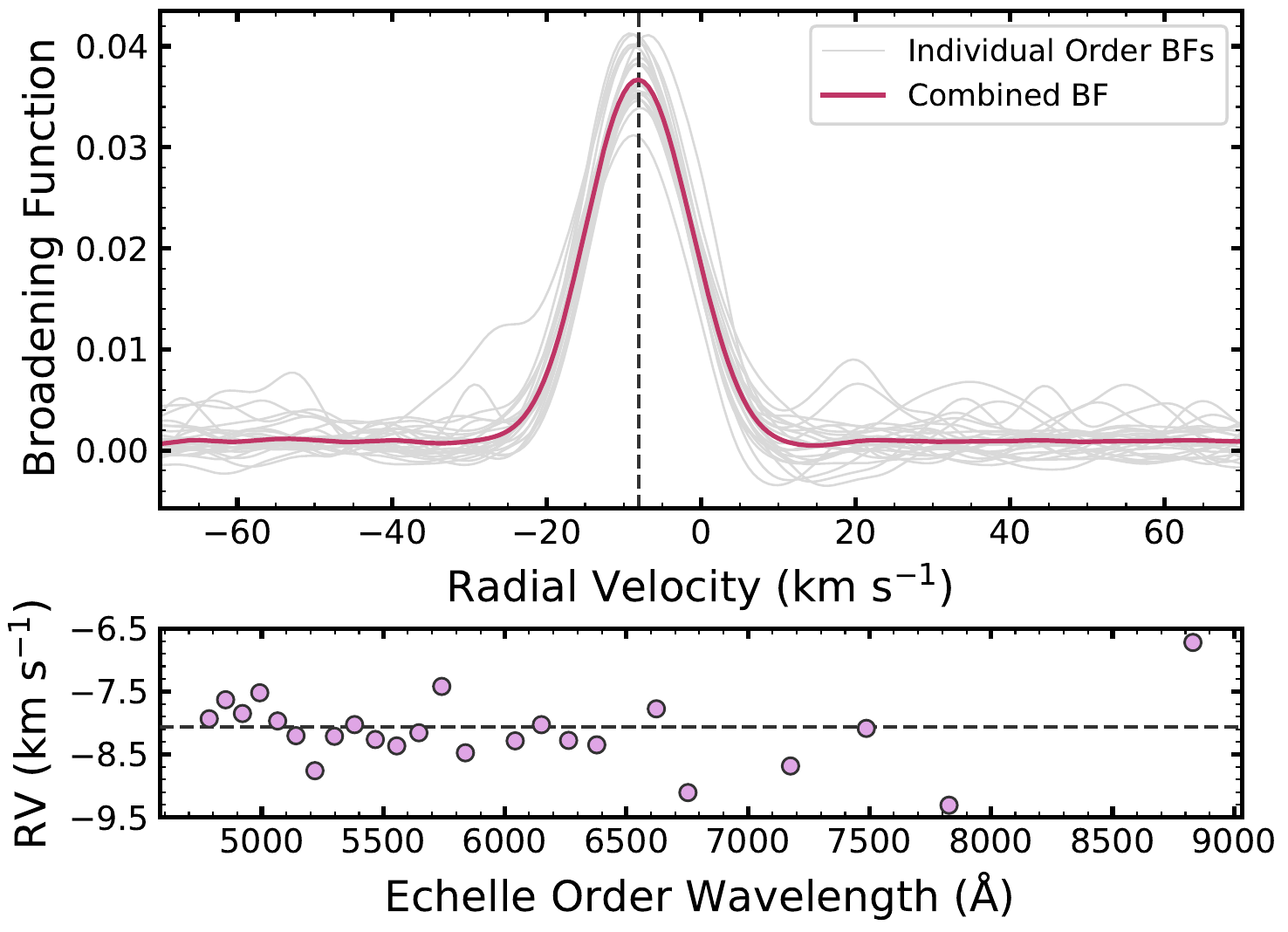}\hfill
\caption{RV measurement using spectra from the Tull coud\'{e} spectrograph at McDonald Observatory, using Gaia EDR3 2280112203742060928 as an example. Top panel: The broadening functions (BFs) computed. The gray lines show the BFs for each individual echelle spectral order included in the RV solution. The red line shows the BF from combining the individual order BFs weighting by their estimated error from the edges of the BF. The combined BF is fit with a Gaussian and used for the final RV measurement. Bottom panel: The RV derived for each order by fitting the individual order BFs with a Gaussian. The individual order RVs largely agree, and feature more scatter at redder wavelengths where there is more telluric contamination.}
\label{fig:RVsol}
\end{figure}

Making use of the {\tt saphires} line broadening functions \citep{Tofflemire19}, we computed radial velocities for each of these stars. A sample RV solution and broadening function is provided in Figure \ref{fig:RVsol}. For the handful of duplicate observations we took the RV measurement with the lowest uncertainty, except in one case where the uncertainties in the measurements were very similar and we took their weighted average. This resulted in 90 sub-km s$^{-1}$ radial velocity measurements spread throughout CFN, with only 10 stars having uncertainties $\sigma_{v_R} > 3$ km s$^{-1}$. These stars with higher uncertainties consistently had severe rotational broadening or internal contamination from a  
spectroscopic binary companion. Two of these stars had sufficiently broad lines that the broadening function fit was quite poor, and these stars are flagged in Table \ref{tab:members}, along with a separate flag for evidence of spectroscopic binarity. These objects are treated as if they have no RVs from this paper, however this does not change any results, as these observed radial velocities are overwritten with superior literature RVs in all cases. In total, our observations resulted in new RV measurements for 113 stars, including 94 from HJST and 19 from LCO. 

\subsection{Youth Indicators}

\begin{figure}
\centering
\includegraphics[width=4.2cm]{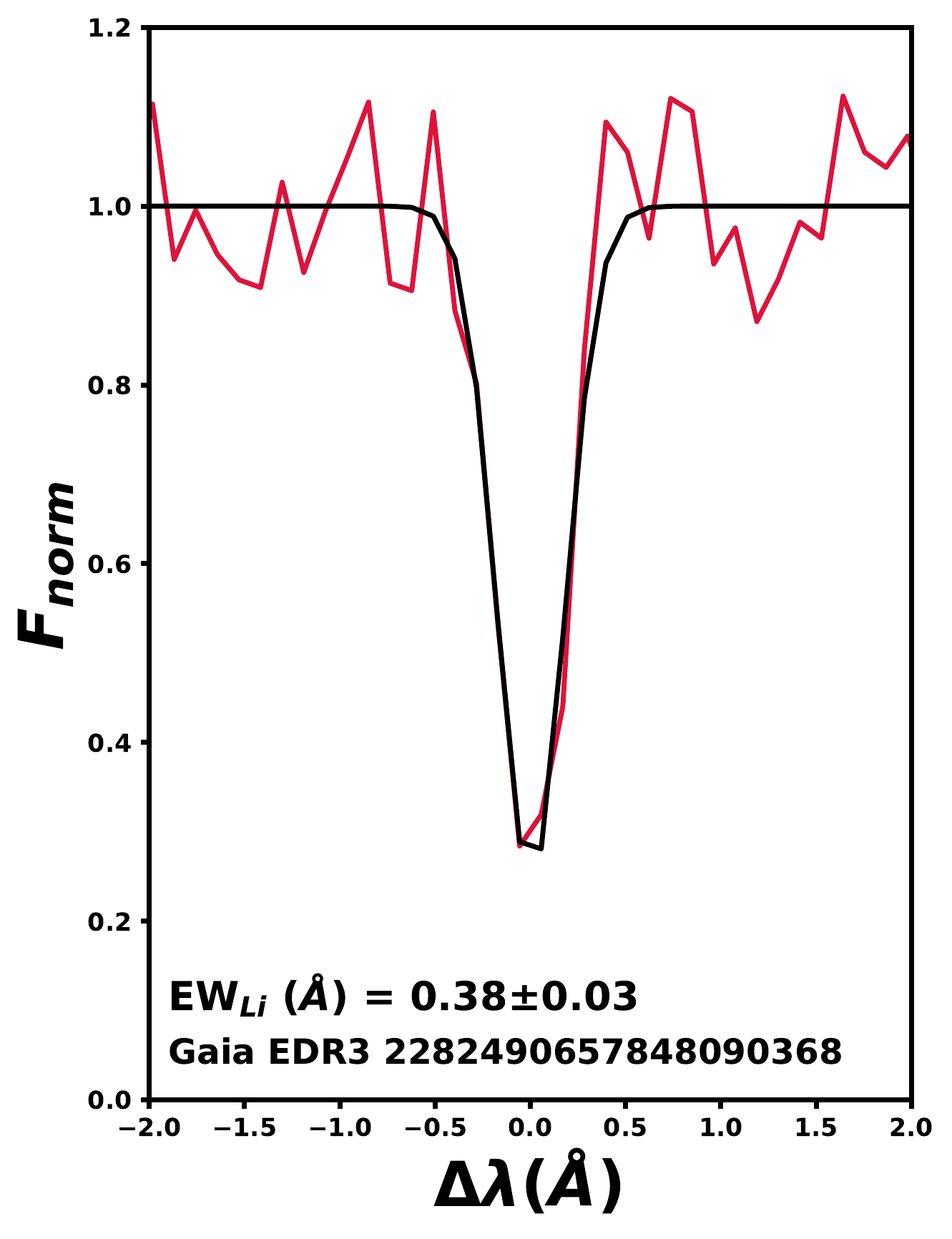}\hfill
\includegraphics[width=4.1cm]{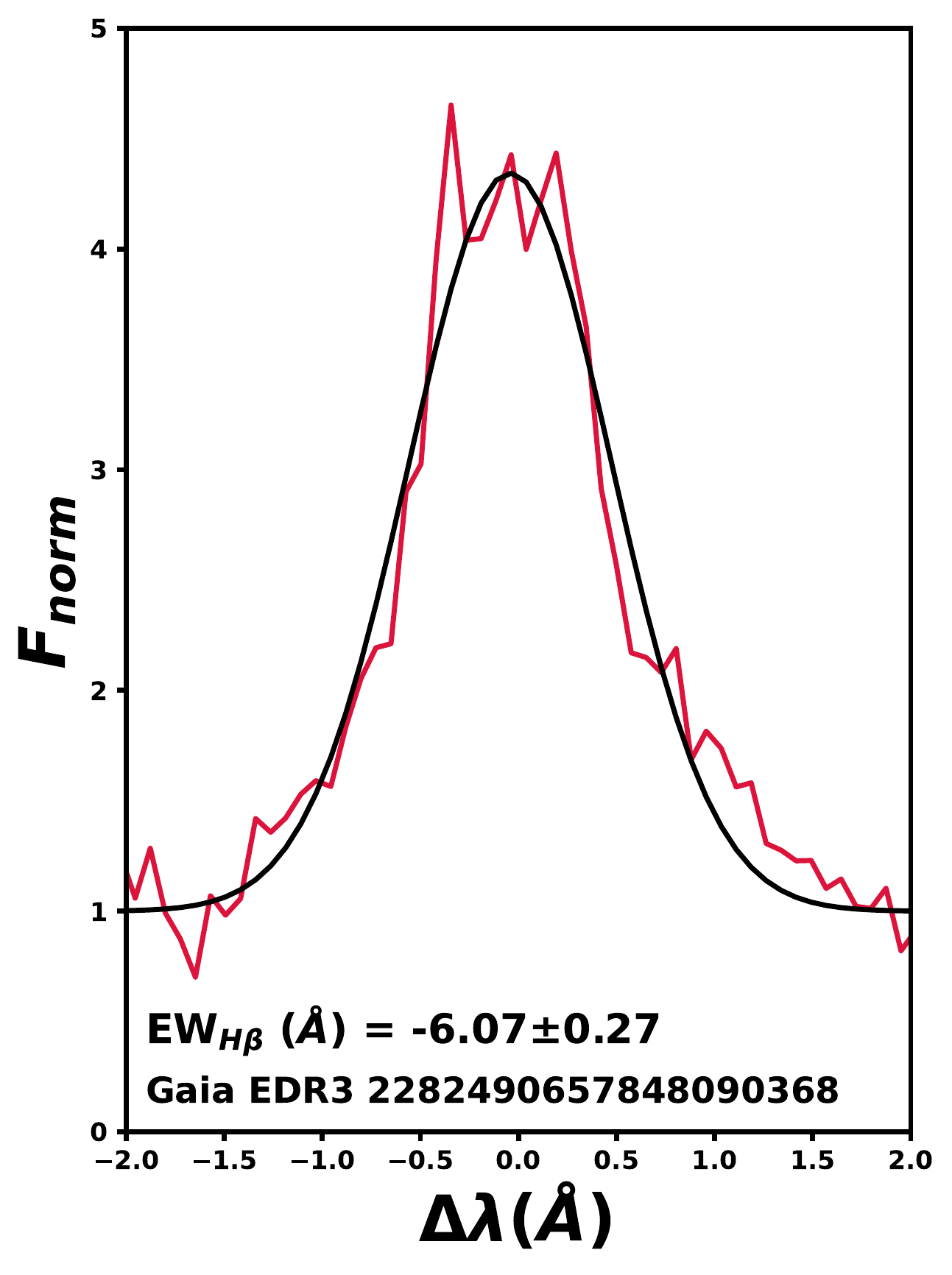}\hfill
\caption{Sample Li and H$\beta$ fits for a typical sample CFN member. The x axis shows wavelength relative to the line centre, as determined from the computed RV measurement and corresponding barycentric correction, and the y axis shows flux, normalized to the continuum flux level. }
\label{fig:LiHbfits}
\end{figure}

Stars that received spectral coverage through LCO or HJST observations can also be checked for youth indicators such as hydrogen emission lines and lithium 6708 \AA~ absorption lines. Due to our limited coverage of the H$\alpha$ line at HJST, we used H$\beta$ in its place, which produces similar emission profiles in young stars, albeit at a lower intensity \citep[e.g.,][]{Frasca10}. Equivalent widths (EWs) for both the Li and H$\beta$ spectral lines were computed by fitting a Gaussian to a limited region around the line in question, using an inverted Gaussian for the Li absorption and an upright Gaussian for H$\beta$ in emission. In both cases, we also fit the background around the line, which we also used to normalize the EW measurements. We do not deblend our Li fits with the nearby 6707.4 \AA~ Fe I lines, which may produce measurement discrepancies on the scale of 10-20 m\AA. However, since our uncertainties are often larger than these expected discrepancies, we expect any impacts on our results to be minimal. Example fits to both lines for an example star in CFN are provided in Figure \ref{fig:LiHbfits}. EW measurements for both Li and H$\beta$ are provided in Table \ref{tab:spectres}. 

\begin{deluxetable*}{ccccccccccc}
\tablecolumns{11}
\tablewidth{0pt}
\tabletypesize{\scriptsize}
\tablecaption{The spectroscopic properties acquired for stars in CFN, including objects observed with HJST and LCO, as well as sources with literature RVs. All Li and H$\beta$ EW measurements are acquired through original observations, while the RV measurements come either from the literature or from observations, with high-quality literature values overwriting our observations in cases of poor RV results.}
\label{tab:spectres}
\tablehead{
\colhead{Gaia ID} &
\colhead{RA} &
\colhead{Dec} &
\multicolumn{3}{c}{RV (km s$^{-1}$)} &
\multicolumn{2}{c}{EW$_{Li}$ (\AA)} &
\multicolumn{2}{c}{EW$_{H\beta}$ (\AA)} &
\colhead{Spectum Source} \\
\colhead{} &
\colhead{(deg)} &
\colhead{(deg)} &
\colhead{val} &
\colhead{err} &
\colhead{src\tablenotemark{a}} &
\colhead{val} &
\colhead{err} &
\colhead{val} &
\colhead{err} &
\colhead{}
}
\startdata
 2280112203742060928 &  329.6421 &  75.0548 &   -8.06 &    0.15 &                 HJST &  0.235 &   0.007 &   0.00 &     0.01 &      HJST \\
 2279209474632504960 &  329.5121 &  73.9760 &  -13.47 &    2.13 &        I/355/gaiadr3 &        &         &        &          &           \\
 2276714476589258752 &  317.1624 &  73.3523 &   -8.40 &    0.08 &                 HJST &  0.413 &   0.033 &   0.00 &     0.00 &      HJST \\
 2276766909550114432 &  317.6483 &  73.8806 &   -8.02 &    0.33 &                 HJST &  0.000 &   0.000 &   4.75 &     0.48 &      HJST \\
 2277005533636485376 &  315.2678 &  74.2236 &  -10.77 &    0.10 &                 HJST &  0.314 &   0.007 &   0.00 &     0.00 &      HJST \\
 2278401986420286336 &  317.8724 &  76.2414 &   -9.94 &    2.61 &        I/355/gaiadr3 &        &         &        &          &           \\
 2278408308612145408 &  318.0061 &  76.3079 &   -7.31 &    1.96 &              LCO-ELP &  0.295 &   0.188 &   0.00 &     0.00 &   LCO-ELP \\
 2278408411691360768 &  318.0164 &  76.3102 &   -6.26 &    1.06 &              LCO-ELP &  0.002 &   0.005 &   0.00 &     0.00 &   LCO-ELP \\
 2279483386169937152 &  335.7461 &  74.7225 &   -8.53 &    0.59 &                 HJST &  0.523 &   0.052 &   0.55 &     0.11 &      HJST \\
 2281351117125711488 &  348.6550 &  77.3510 &   -7.57 &    0.50 &                 HJST &  0.413 &   0.040 &   3.07 &     0.18 &      HJST \\
 2278683461396881152 &  315.5097 &  76.9709 &   -3.91 &    0.68 &        I/355/gaiadr3 &        &         &        &          &           \\
 2277380677556833920 &  321.9343 &  75.4817 &   -8.19 &    0.38 &                 HJST &  0.428 &   0.043 &   2.12 &     0.21 &      HJST \\
\enddata
\tablenotetext{a}{The source of the RV measurement: either the observatory source of our new observation or the ID of the vizier table where it can be found}
\vspace*{0.1in}
\end{deluxetable*}

\section{Membership} \label{sec:membership}

Our selection of CFN members has two separate goals: first, to identify candidate members to help us assess the association's stellar populations and substructure, and second, to identify a robust set of well-behaved high-confidence members for use in  kinematic traceback and other mass estimation methods. For a complete census of the association, only near-certain non-members should be removed to allow even the most tenuous possible members to be retained for additional testing. Similarly permissive restrictions are also desirable for finding substructure, as the HDBSCAN clustering algorithm is designed for cluster identification against a relatively uniform background, and a larger sample invariably means more stars to help resolve more tenuous subgroups. Our kinematic traceback sample must be more restrictive, however, as the inclusion of non-members would greatly dilute the motions of genuine members, as would including stars with inconsistent or even unreliable velocities. We have already largely defined a set of credible targets through our initial space-velocity candidate selection in Section \ref{sec:candidateselec}, however there are many other ways in which we can restrict the sample, including cuts on photometry, stellar motions, and binaries. In this section, we describe the cuts we use, and how we use them to produce reliable CFN candidate lists. 

\subsection{Photometric Selection} \label{sec:photselec}

Pre-main sequence stars have notably elevated luminosities compared to typical field stars of similar $T_{\rm eff}$, which allowed SPYGLASS-I to identify numerous young associations such as CFN. However, now that we have verified the existence of CFN and know the position-velocity parameter space it covers, we can revise the region of the CMD in which youth can be confidently asserted. Unlike our detection of CFN in SPYGLASS-I, where we were attempting to pull subtle features out of a dominant field, here we separate field contamination from a known population while the two exist in relatively similar numbers. This environment with a more dominant young population greatly increases the probability that a star located closer to the zero-age main sequence is a genuine cluster member rather than an older photometric impostor such as a field binary. 

To reassess the probability that stars are CFN members, we use a modified version of the SPYGLASS Bayesian approach. Instead of comparing the probability of youth within a 10 million system field model, we instead generate separate models for the typical SPYGLASS young populations and the broader field, and compare the probability of membership in each model. As we later show in Figure \ref{fig:Mem_CMD}, the field main sequence and CFN pre-main sequence are quite well-separated over much of the CMD, especially for intermediate masses with $7 \la M_G \la 9$. We use this well-separated region to set the priors on the relative sizes of the young and field populations, counting the young versus old stars that lie between the 0.5 and 0.8 M$_{\odot}$ solar-metallicity isomass tracks from the PARSEC v1.2S isochrone models \citep{PadovaBressan12}. We used the PARSEC v1.2S 50 Myr solar metallicity isochrone as a concrete definition of the dividing line between these younger and older populations, a choice which cleanly bisects the two while mirroring our minimum age for consideration as a young population in SPYGLASS-I. We found that 53$\pm$5\% of stars between the 0.5 and 0.8 M$_{\odot}$ isomasses lie above that 50 Myr isochrone and therefore appear likely young, with uncertainties added based on binomial statistics. To reflect this percentage we assembled a model consisting of 4.7 million field systems and 5.3 million systems drawn from properties consistent with young populations recognized through SPYGLASS.

The field model we use is identical to the one described in SPYGLASS-I, drawing primary stellar properties from the \citet{Chabrier05} individual object Initial Mass Function, the GALAH DR2 metallicity distributions \citep{Hayden19}, a uniform age probability distribution below the 11.2 Gyr solar neighborhood age from \citet{Binney00}, and generating multiple companions derived from the binarity rates from \citet{Duchene13}, the mass ratio distributions from \citet{Kraus11}, and system separations from \citet{Raghavan10}. A much more detailed description of our field model generation can be found in SPYGLASS-I. 

The model we generate to represent a standard SPYGLASS young stellar population is similar to our field model, but limits stars to ages below 50 Myr, and draws from a more limited set of metallicities drawn from a normal distribution centered on solar metallicity with $\sigma=0.1$. This reflects the typically solar metallicities known in nearby young associations, as most nearby associations have metallicities within $< 1 \sigma$ \citep[e.g.][]{Almeida09}, and the more non-solar-metallicity open clusters, such as M35 ([Fe/H]=-0.21; \citealt{Bouy15}) and Praesepe ([Fe/H]=+0.21; \citealt{Cummings17}) have metallicities discrepant at no more than $\sim 2 \sigma$. The merging of photometry in close multiple systems for both the field and young populations is calibrated to 179 pc, the mean distance to CFN recorded in SPYGLASS-I. 

Using the same Bayesian formula from SPYGLASS-I, we compute the probability of membership in the field population versus the model young population. Like in SPYGLASS-I, we use a corrective prior equal to the age of the model star for the field population to maintain a uniform age distribution despite the uniform log-space sampling which we used to generate model stars. The uniform log-space sampling was chosen to improve our coverage of the pre-main sequence where stellar evolution progresses more rapidly than on the main sequence, creating the sampling imbalance corrected by this prior. We then set the prior on the young member model equal to the mean prior on the field model, which sets the sum of the young member priors at a 53\% of the model's total, consistent with the share of the total model occupied by our young member model. Cuts produced through the resulting probabilities visibly separate the field sequence from the pre-main sequence of CFN, especially lower on the PMS where we are not able to provide radial velocity follow-up. We use these probabilities in Section \ref{sec:sss} for selecting a sample of robust CFN candidate members, which is used for much of our later analysis. 

\subsection{Velocity Selection} \label{sec:rv_selec}

Having a complete 3D velocity vector consistent with the association serves as a strong indication for membership, and the presence of a radial velocity measurement is critical to completing that vector. These velocities form a metric fully independent of the prior selection steps, providing a highly complementary indicator of association membership. The sensitivity of LCO and HJST allows for the collection of high-resolution spectra of all CFN candidate members above the apparent divergence point between the pre-main sequence of CFN and main sequence field interlopers at $G_{BP}-G_{RP} \la 1.4$, so RV follow-up is complementary in establishing membership alongside photometric methods.

While our RV coverage in CFN is essentially complete among the more massive members of the association, it remains low relative to the overall population, with a selection that includes very few probable non-members. As such, our RV data is not well-suited for understanding the prevalence of background interlopers, a factor critical to selecting stars with reliably consistent RVs, be it through the re-definition of CFN using radial velocities to enable HDBSCAN clustering in 6D space-velocity coordinates, or through any comparable approach that relies on identifying an RV interval that is overdense relative to the background velocity distribution. We therefore instead approximate the velocity distribution of CFN as spherical in UVW space, with a maximum radius equal to the maximum projected radius of the transverse velocity distribution, which we use to set our threshold. We measure this maximum projected radius as 7 km s$^{-1}$ relative to the median transverse velocity. Applying this to our velocities in UVW space, we define a cut that preserves stars within a radius of 7 km s$^{-1}$ of our median UVW value, \edit1{(U,V,W) = (-10.08, -12.31, -5.31).} 

\begin{figure}
\centering
\includegraphics[width=8.2cm]{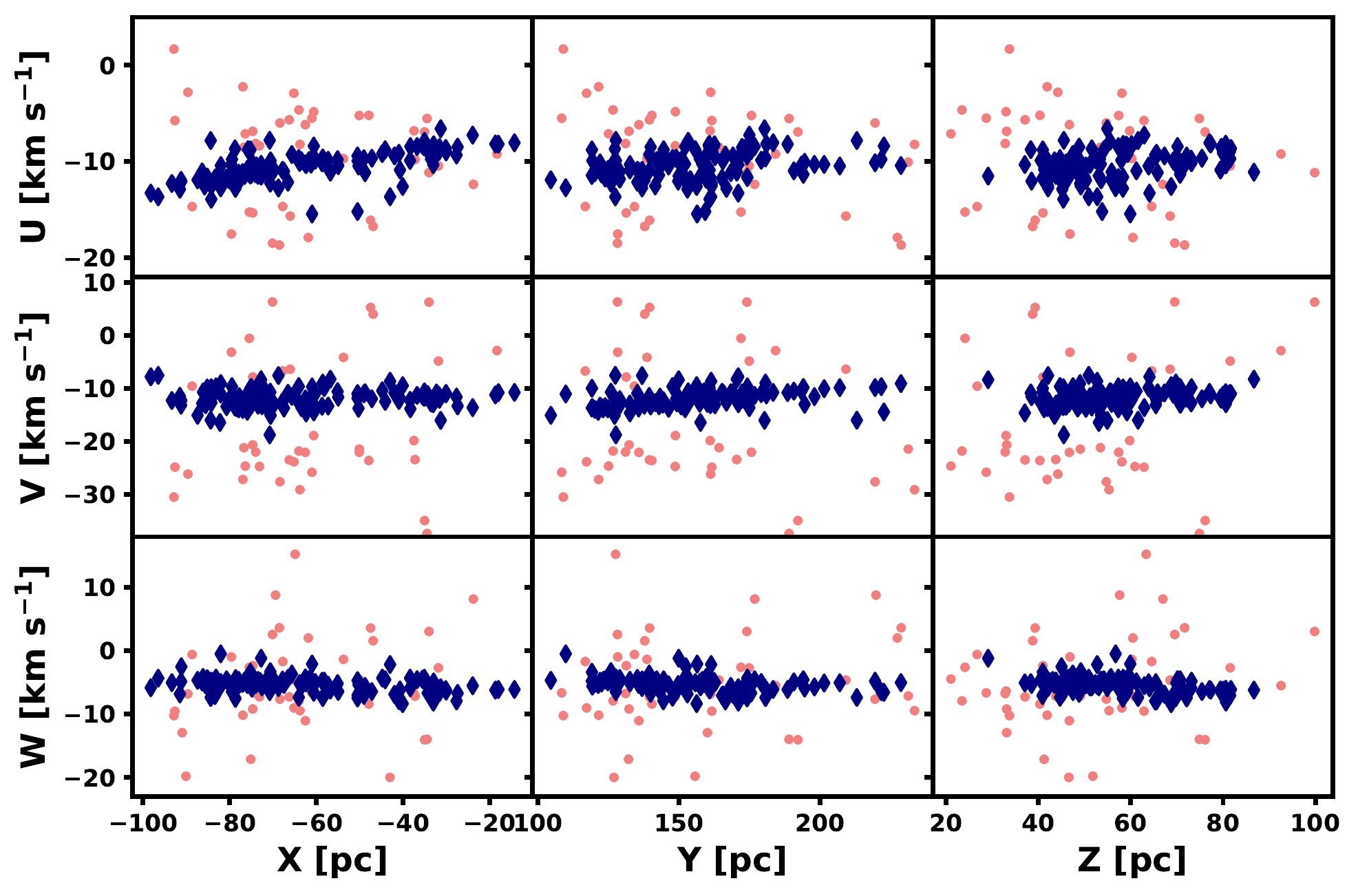}\hfill
\caption{Distribution of the three velocity axes, plotted against the three spatial axes. Stars marked as dark blue diamonds pass our RV cut, while stars marked by light red dots do not.}
\label{fig:UVWcuts}
\end{figure}

The result of this velocity cut is presented in Figure \ref{fig:UVWcuts}, which shows how each velocity component varies in spatial coordinates. While there is some spatial dependence to the UVW velocities, the amplitude of these variations is less than the magnitude of this cut, indicating our single-cut approach is sufficient for restricting the sample without a risk of significant loss of genuine members. Furthermore, the radial velocity measurements that survive the cut follow the space-velocity trends visible in Figure  \ref{fig:UVWcuts} quite reliably, as expected for an internally coherent association. We describe our application of this cut to our overall sample in Section \ref{sec:sss}. 

\subsection{Binaries} \label{sec:binaries}

The presence of binaries significantly complicates the reliable identification of stars as CFN members, as these objects often show errant motions arising from the influence of the companion \citep[e.g.,][]{Offner22}. While binary systems tend to be well-behaved in spatial coordinates, these high internal velocities significantly raise their distance from other members in the 5-dimensional space-velocity parameter space we used for clustering and the identification of candidate members. This makes binaries probable outliers in velocity space, by extension increasing the probability that they will be missed when searching for CFN candidate members and reducing their utility for dynamical studies.

To locate stars with a binary companion influencing their kinematics, we perform a search of the Gaia EDR3 catalog for each member that would survive our broadest photometric-only cut defined in Section \ref{sec:sss}, scanning a region within 10000 AU of the star at the distance recorded in Gaia EDR3. While binaries wider than this are known to exist, they are relatively rare, representing less than 4\% of the total population \citep{Raghavan10,Offner22}. Furthermore, these ultra-wide binaries have orbital velocities low enough that they are unlikely to produce velocity dispersions capable of influencing dynamical studies. All objects within this search radius with parallaxes with $\frac{\Delta \pi}{\pi} < 0.2$ and proper motions with $\Delta \mu < 5$ mas yr$^{-1}$ were identified as likely companions, with those cuts chosen to provide generous restrictions given typical uncertainties and expected internal motions within binary systems, respectively. The resulting search produced 163 stars that are a component of a binary or multi-star system, with 58 of these systems consisting purely of members in our CFN likely member list, and the remaining 21 consisting of likely member and one member not in that sample. Most of the binary components outside of our likely member sample were in the set of stars with $D<0.05$ that we removed early in our sample selection. Members of binary systems in our CFN likely member sample are flagged in Table \ref{tab:members}, and we provide a table containing all identified pairs in Appendix \ref{app:bin} for use in future works.

Unresolved binaries must also be excluded from our sample, as unidentified companions can skew RVs, interfere with the background flux for lithium Depletion measurements, and lower the apparent isochronal age of the star. A visible double-line broadening profile in spectroscopic observations is the clearest indication of this, however this is only a useful metric where we have collected observations. We can also identify cases where stars are resolved in Gaia, but too close to get a clean spectroscopic measurement of a single star through spectral observations. For all other situations, the Renormalized Unit Weight Error, or RUWE, is a frequently used metric for identifying unresolved binaries, as it effectively quantifies the deviation from a quality single-source astrometric model that is often induced by such unresolved sources. Following \citet{Bryson20}, RUWE$>$1.2 is dominated by binaries, and the contamination from binaries becomes very small for RUWE$<$1.1, so both of these cuts have uses for different contamination tolerances. \citet{Fitton22} recently showed that young stars often have significantly higher RUWEs than older stars, especially when they also host disks, with the 95th percentile for single stars at RUWE$=1.6$ without disks and RUWE$=2.5$ with them. While we recognize that these inflated RUWEs are likely to produce increased losses among single stars when setting RUWE cuts for binaries the overall parameter space occupied by binaries appears largely unchanged, making a cuts at RUWE$>$1.2 still appropriate for removing most binaries, and RUWE$>$1.1 still appropriate when an essentially clean data set is required. Our precise application of each condition is described in Section \ref{sec:sss}. Experimentation with these cuts consistently shows higher luminosities among stars exceeding these RUWE limits, which supports the conclusion that most of these objects are binaries.

\begin{deluxetable*}{ccccccccccccccc}
\tablecolumns{15}
\tablewidth{0pt}
\tabletypesize{\scriptsize}
\tablecaption{All stars identified as credible candidate members of the Cepheus Far North association, including basic properties from Gaia, their parent subgroup, their estimated mass, the presence of a disk, and various quality flags. We include flags for photometric membership, velocity membership, and binarity, which are used to produce the higher-confidence samples used in later analyses (see Section \ref{sec:sss}).}
\label{tab:members}
\tablehead{
\colhead{Gaia ID} &
\colhead{SG\tablenotemark{a}} &
\colhead{RA} &
\colhead{Dec} &
\colhead{$m_G$} &
\colhead{$G_{BP}-G_{RP}$} &
\colhead{$\pi$} &
\colhead{M} &
\colhead{$disk?$} &
\colhead{$D$} &
\colhead{$A$\tablenotemark{b}} &
\colhead{$P$\tablenotemark{c}} &
\colhead{$V$\tablenotemark{d}} &
\colhead{$PY$\tablenotemark{e}} &
\colhead{$F$\tablenotemark{f}} \\
\colhead{} &
\colhead{} &
\colhead{(deg)} &
\colhead{(deg)} &
\colhead{} &
\colhead{} &
\colhead{(mas)} &
\colhead{(M$_{\odot}$)} &
\colhead{} &
\colhead{} &
\colhead{} &
\colhead{} &
\colhead{} &
\colhead{} &
\colhead{}
}
\startdata
 2276643622512797440 &       2 &  314.0388 &  72.9399 &  17.42 &  3.25 &      4.83 &  0.22 &      0 &  0.11 &  1 &  1 &  0 &   1 &   0 \\
 2280112203742060928 &       1 &  329.6421 &  75.0548 &  10.20 &  0.98 &      5.85 &  1.32 &      0 &  0.31 &  1 &  1 &  1 &   1 &   9 \\
 2280112208035806336 &       1 &  329.6448 &  75.0568 &  16.21 &  2.89 &      5.89 &  0.31 &      0 &  0.30 &  1 &  0 &  0 &   1 &   1 \\
 2279209474632504960 &       2 &  329.5121 &  73.9760 &  12.85 &  1.09 &      4.22 &  0.84 &      0 &  0.15 &  1 &  1 &  1 &   0 &   0 \\
 2276697365440396032 &       5 &  314.2924 &  73.4494 &  14.71 &  2.65 &      5.41 &  0.46 &      0 &  0.25 &  1 &  1 &  0 &   1 &   8 \\
 2276697571598823424 &       2 &  314.2421 &  73.4652 &  15.03 &  2.72 &      4.90 &  0.43 &      0 &  0.10 &  0 &  1 &  0 &   1 &   8 \\
 2276714476589258752 &       5 &  317.1624 &  73.3523 &  12.48 &  1.30 &      5.06 &  0.85 &      0 &  0.13 &  1 &  1 &  1 &   0 &   0 \\
 2276726674296374656 &       1 &  317.5639 &  73.6021 &  15.93 &  2.62 &      5.95 &  0.40 &      0 &  0.11 &  1 &  1 &  0 &   0 &   0 \\
 2276736840483401856 &       6 &  316.9934 &  73.5380 &  18.37 &  3.65 &      5.90 &  0.15 &      0 &  0.14 &  1 &  1 &  0 &   0 &   0 \\
 2276738077434022016 &       6 &  316.9527 &  73.5930 &  18.17 &  4.05 &      6.16 &  0.12 &      0 &  0.12 &  1 &  1 &  0 &   1 &   1 \\
 2276738081729485696 &       6 &  316.9391 &  73.5925 &  14.86 &  2.62 &      6.01 &  0.47 &      0 &  0.10 &  1 &  1 &  0 &   1 &   9 \\
 2276746602944652800 &       2 &  318.1262 &  73.6225 &  18.46 &  3.79 &      3.96 &  0.15 &      0 &  0.08 &  1 &  1 &  0 &   1 &   0 \\
\enddata
\tablenotetext{a}{The ID of the subgroup the star is assigned to}
\tablenotetext{b}{The boolean solution to the Astrometric quality cut from SPYGLASS-I, which is based on the unit weight error. 1 passes, 0 fails. }
\tablenotetext{c}{The boolean solution to the Photometric quality cut from SPYGLASS-I, which is based on the BP/RP flux excess factor. 1 passes, 0 fails.}
\tablenotetext{d}{A flag to represent the results of our velocity membership cuts. 1 passes, -1 fails, and 0 has no RV.}
\tablenotetext{e}{A flag to represent our photometric membership calculation. A value of 1 marks stars with $\Sigma P_{mem}$ greater than 10\% of the total, which contains 90\% of genuine members and excludes most outliers, and 0 has photometry that is neither conclusively non-young nor likely young. Stars with failed photometric flags (-1) are not included in this table.}
\tablenotetext{f}{A flag for other notable features. 1 indicates that the star has a resolved companion within 10000 AU in the plane of the sky, 2 indicates a bad broadening function solution, 4 indicates a bimodal line profile likely indicative of spectroscopic binarity, 8 indicates an RUWE$>$1.2, indicating likely unresolved binarity, and 16 indicates that the RV recorded was ambiguously attributed to two components of a binary pair. The flags are added in cases where multiple are true; for example, flag 6 indicates both flag 2 and 4. }
\vspace*{0.1in}
\end{deluxetable*}

\subsection{Final Stellar Selections} \label{sec:sss}

For our demographic and substructure-defining sample, we select a very lenient cut with the goal of removing only near-certain non-members, which allows more marginal sources with good RVs to be confirmed as members using velocity measurements prior to the imposition of harsher photometric membership cuts. We therefore apply a cut such that $\Sigma P_{mem} < 1$  for all excluded candidates, a selection which implies that only one genuine member will be excluded. Stars that fail this cut are removed from all subsequent analysis. \edit1{Three} stars that failed this cut were observed at HJST \edit1{or LCO, and none of them showed conclusive spectroscopic youth indicators}, however we include them in Table \ref{tab:spectres}, the spectroscopic data table, for the sake of completeness. This cut rejected 352 objects, leaving 549 photometrically credible candidates, and 552 total objects \edit1{covered across Tables \ref{tab:spectres} and \ref{tab:members} including the three }non-members covered by HJST spectroscopic observations. We do not apply any RV cuts to this broader set, as members of multiple systems will often have velocities significantly different from that of the system's barycenter, making it difficult to confidently reject a member on purely kinematic grounds. 

\begin{figure}
\centering
\includegraphics[width=8.5cm]{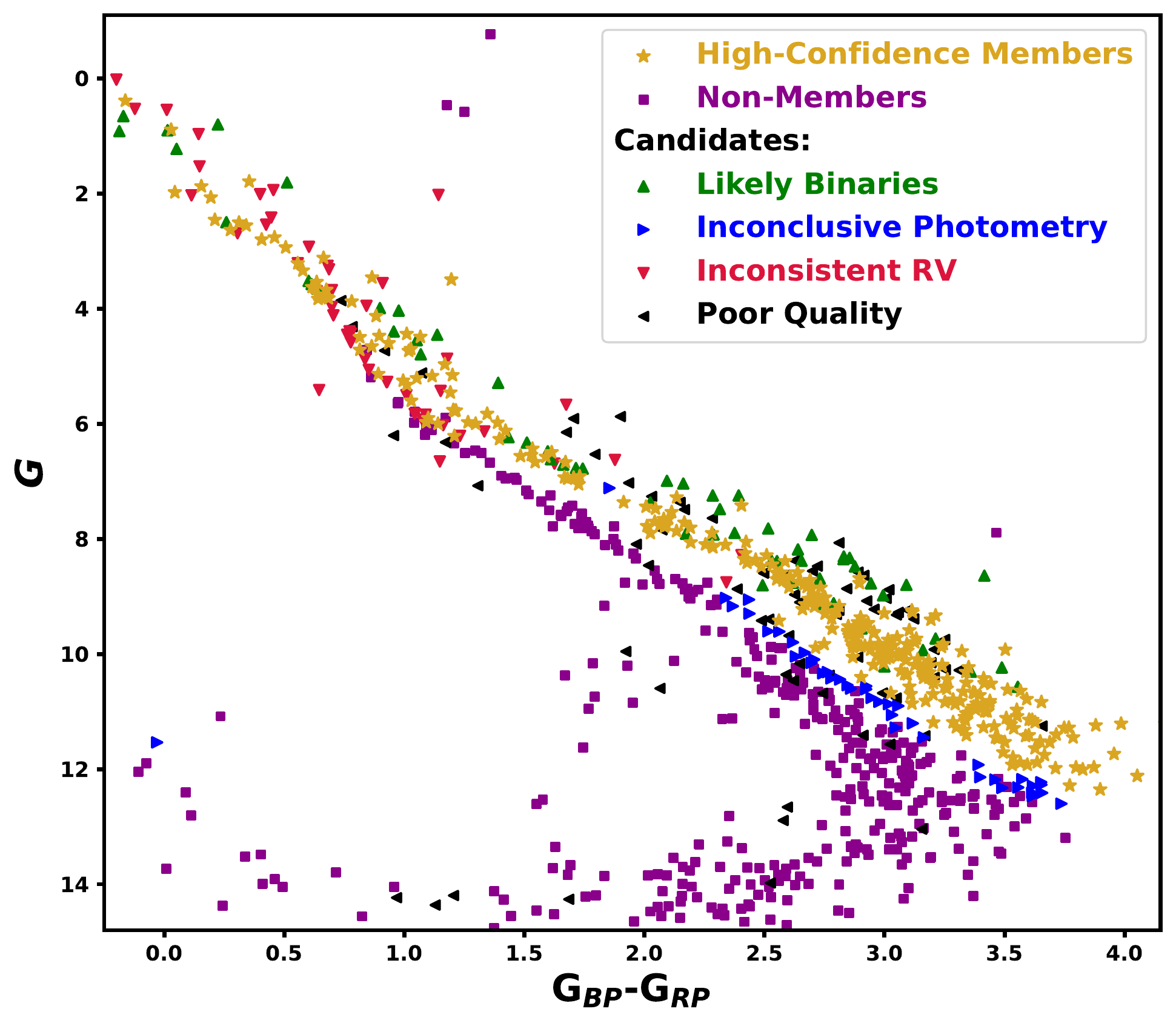}\hfill
\caption{CMD of stars in our CFN sample. Yellow stars indicate objects that pass all restrictions. Purple squares are removed by our hard photometric cut \edit1{and are therefore considered non-members}, red inverted triangles are removed by our velocity cut, black left-pointing triangles are removed due to photometric or astrometric quality cuts, and blue right-pointing arrows have no radial velocities and are rejected due to the soft photometric cut. Green upward-pointing arrows pass all other cuts, but are rejected by our RUWE cut and our accompanying spectroscopic binarity checks, comprising primarily stars elevated above the pre-main sequence by the presence of an unresolved companion.}
\label{fig:Mem_CMD}
\end{figure}

For our set of high-confidence members used as the basis for kinematic traceback analysis, isochronal ages, and lithium depletion studies, we produce a more restrictive sample that introduces our velocity cuts while further considering the photometric membership probabilities defined in Section \ref{sec:photselec}. The goal is to produce a sample consisting of all high-quality single stars, provided that they have a combination of photometry and velocity consistent with youth. We start by removing stars with velocities inconsistent with the association by applying the UVW cut described in Section \ref{sec:rv_selec}, which removes stars more than about 7 km s$^{-1}$ from the median UVW value. This cut excludes a few stars with high youth probabilities, mainly towards the upper end of the CMD, typically due to binarity significantly altering the velocities relative to a comoving barycenter. While many stars that fail such a cut may be legitimate members, especially binaries with significant internal motion, they will not be useful for traceback due to the discrepancy between the measured RV of a single component and the barycentric motion of the system, and contamination from close companions is likely given the high velocities induced on these stars.

With these RV-inconsistent candidates removed, we provide two routes through which stars remain in our sample: one through velocity for stars where 3D data is available, and one through photometry, which is mainly used for lower-mass stars high on the pre-main sequence. For the velocity condition, we include all stars with velocity within that 7 km s$^{-1}$ cut from Section \ref{sec:rv_selec}, as long as they pass the Gaia astrometric quality cut (see Section \ref{sec:candidateselec}). For our photometric condition, we include stars that pass a cut on our photometric membership probabilities such that $\Sigma P_{mem}$ of the excluded members is less than 10\% of the total $\Sigma P_{mem}$, as long as the stars do not fail our photometric or astrometric quality cuts. The latter condition is designed to retain 90\% of genuine members, and while over 160 stars fail that $P_{mem}$ cut, only about 40 stars are actually removed by it due to many of these stars resting higher on the main sequence where they are confirmed using radial velocity membership condition. The result of our observational design is that more massive stars typically have RV coverage that prevents their removal through the photometric conditions by virtue of their ambiguous photometric age, while less massive stars can be reliably vetted through their photometry. 

Finally, we remove all sources with a single RV measurement assigned to multiple sources, sources with visible spectroscopic double-line profiles, and objects with RUWE$>1.2$. All of these cuts help to remove likely unresolved binaries from the sample. The RUWE cut chosen is the softer of the two cuts proposed in Section \ref{sec:binaries}, and we apply it here to ensure a sample mostly clear of binaries, but not so much so that the overall sample size suffers considerably. Further RUWE cuts are useful for isochrone fitting due to the larger sets of available data encouraging sample purity over completeness, however we apply them prior to the isochrone fitting and not for this broader sample. We take a similar approach with resolved binary companions. Any source with a companion that can induce motion must be removed for accurate dynamical studies, however such a companion should have little impact on measurements of Li or photometry, so we only remove these wide companions when performing our dynamical studies. 
The combination of these restrictions essentially ensures that all single stars that pass quality flags and have feasibly young photometry and RVs consistent with CFN are included in later analyses, along with high-confidence photometrically young stars provided that they are not excluded from consideration by inconsistent RVs. The resulting sample contains 302 stars.

\begin{figure}
\centering
\includegraphics[width=7.5cm]{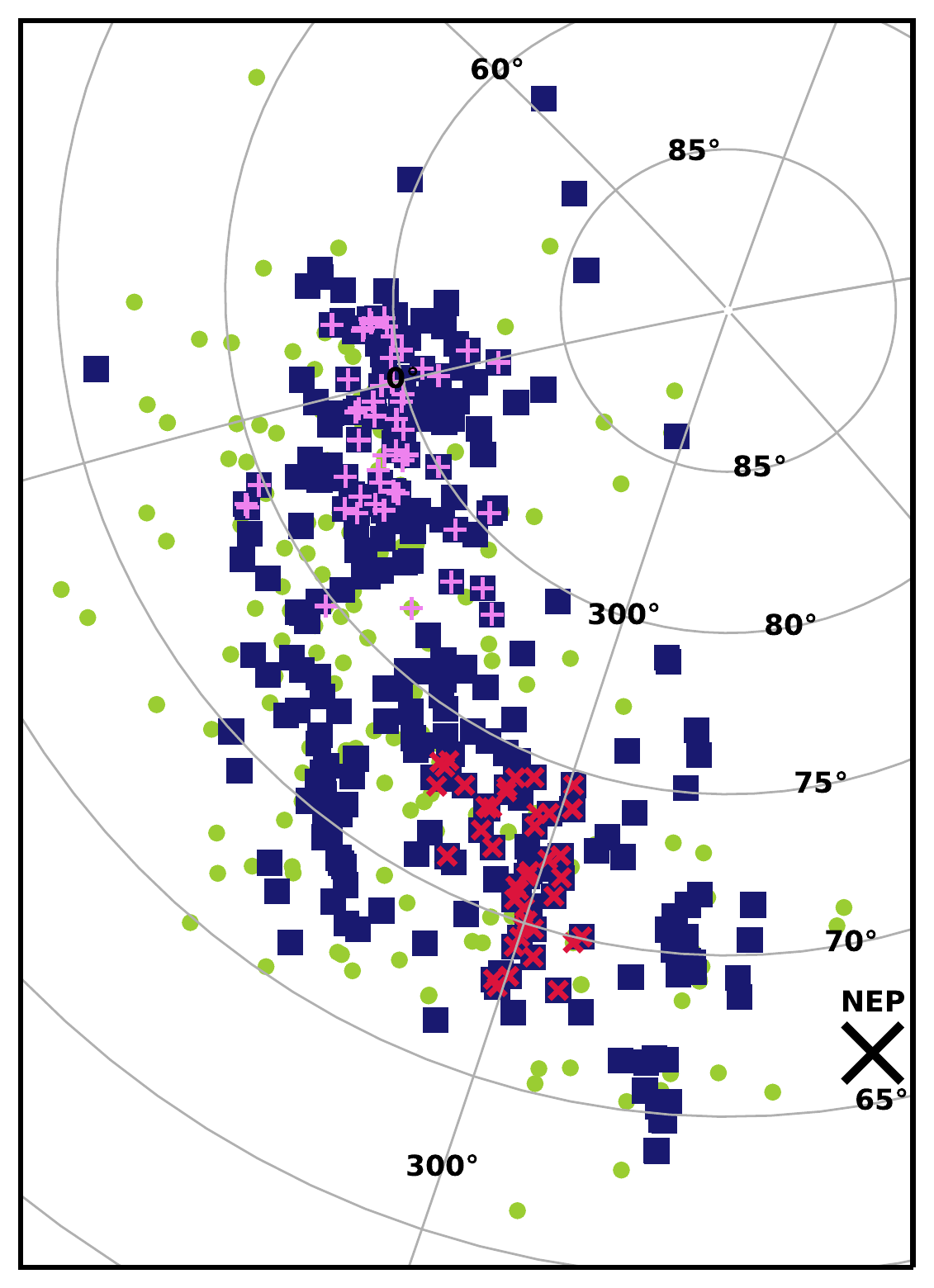}\hfill
\caption{Stars in CFN, presented in RA/Dec sky coordinates. Blue squares indicate stars in the high-confidence member sample, while the green circles indicate stars only in the broader sample candidate members. The North Ecliptic Pole is marked for reference. \edit1{Stars originally in the \citet{Szilagyi21} catalogs for the CFN Core (Cepheus Association) and HD 190833 are marked with purple plusses and red x's, respectively.}}
\label{fig:Mem_Sky}
\end{figure}

In Figure \ref{fig:Mem_CMD}, we provide a color-magnitude diagram summarizing stellar selection in CFN, showing stars removed by the photometric cuts, velocity cuts, binary cuts, and general quality cuts, alongside the final high-confidence member sample. Figure \ref{fig:Mem_Sky} displays the population of CFN in RA/Dec sky coordinates, showing stars in the high-confidence stellar sample alongside stars only found in the broader candidate sample. \edit1{We also mark the previous coverage of the region through the \citet{Szilagyi21} catalog, showing the known extent of the association prior to this publication.} The populations we provide in CFN span approximately 30 degrees on the plane of the sky, with sections close to both the north celestial and north ecliptic poles. 

\subsection{Stellar Masses} \label{sec:masses}

The status of CFN and its subgroups as associations depends on the gravitational binding state of its members, as sufficient gravitational binding would classify the group or some localized sub-component within as an open cluster. This has important implications for the dynamical study of the association, as gravitationally bound clusters will not steadily disperse from the time of their formation, making dynamical age estimation much less feasible. We therefore make mass measurements by comparison to solar-metallicity PARSEC isochrones \citep{PARSECChen15}. We generate a set of isomass tracks from those isochrones, which are spaced at every 0.005 M$_\odot$ between 0.09 and 1 M$_\odot$, every 0.01 M$_\odot$ between 1 and 2 M$_\odot$, and every 0.02 M$_\odot$ between 2 and 4 M$_\odot$. We assign the mass of the nearest model track to each star. The masses we produce broadly align with the initial mass function from \citet{Chabrier05}, albeit with a deficiency for stars with M$<0.2$ M$_{\odot}$ likely caused by lower survey sensitivity there. 

\section{Results} \label{sec:results}

\subsection{Substructure} \label{sec:substructure}

\begin{deluxetable*}{ccccccccccccccccc}
\tablecolumns{17}
\tablewidth{0pt}
\tabletypesize{\scriptsize}
\tablecaption{Subgroups we identify in CFN, including the parent node the subgroup formed out of (either EE Dra (EED) or $\beta$ Cep (BCP); see \S \ref{sec:traceback}), and their general properties. N provides the number of credible members out of the total sample of 549 reside in the subgroup, while RA, Dec, distance, and X/Y/Z galactic coordinates are median values, and use the more restricted set of 302 probable members to minimize contamination in these bulk values. The U/V/W velocities are also median values, and we restrict the sample further to require no resolved binaries to avoid the result being inflated by companions. The value of $\sigma_{1D}$ is based on transverse motions to improve the sample size relative to a three-dimensional data set, which requires RVs. The age given is the adopted age, which is also shown in Table \ref{tab:ages}, alongside alternative age solutions. }
\label{tab:sgprops}
\tablehead{
\colhead{SG} &
\colhead{Node} &
\colhead{Name} &
\colhead{N} &
\colhead{M$_{tot}$} &
\colhead{RA} &
\colhead{Dec} &
 \colhead{$d$} &
 \colhead{X} &
 \colhead{Y} &
 \colhead{Z} &
 \colhead{r$_{hm}$ \tablenotemark{a}} &
 \colhead{U} &
 \colhead{V} &
 \colhead{W} &
 \colhead{$\sigma_{1D}$} &
\colhead{Age} \\
\colhead{} &
\colhead{} &
\colhead{} &
\colhead{} &
\colhead{(M$_{\odot}$)} &
\colhead{(deg)} &
\colhead{(deg)} &
\colhead{(pc)} &
\colhead{(pc)} &
\colhead{(pc)} &
\colhead{(pc)} &
\colhead{(pc)} &
\colhead{(km s$^{-1}$)} &
\colhead{(km s$^{-1}$)} &
\colhead{(km s$^{-1}$)} &
\colhead{(km s$^{-1}$)} &
\colhead{(Myr)}
}
\startdata
  1 & BCP &             & 172 &  71.3 &  336.8 &  79.5 &  156.4 & -73.6 &  128.9 &  47.0 &  12.26 & -11.21 & -13.28 & -4.85 &    0.63 &         17.8 $\pm$ 0.8 \\
  2 & BCP & $\beta$ Cep &  95 &  46.1 &  320.9 &  71.4 &  227.9 & -68.6 &  209.7 &  58.2 &  17.70 & -10.44 &  -9.97 & -5.03 &    0.68 &         22.7 $\pm$ 1.3 \\
  3 & EED & EE Dra      &  35 &  16.4 &  285.2 &  69.7 &  192.6 & -32.2 &  172.2 &  80.0 &   2.37 &  -9.15 & -11.17 & -6.26 &    0.28 &         25.8 $\pm$ 2.7 \\
  4 & BCP &             &  75 &  35.2 &  347.4 &  77.1 &  187.4 & -85.2 &  160.0 &  50.1 &  12.74 & -11.88 & -12.19 & -5.02 &    0.88 &         16.0 $\pm$ 1.5 \\
  5 & BCP &             &  94 &  42.4 &  306.2 &  72.0 &  175.6 & -43.3 &  160.2 &  56.8 &  16.14 &  -8.77 & -11.67 & -4.84 &    1.08 &         19.4 $\pm$ 1.3 \\
  6 & EED &             &  52 &  16.6 &  297.9 &  73.5 &  184.3 & -46.6 &  165.4 &  69.2 &  13.72 &  -9.56 & -11.08 & -7.01 &    0.35 &         22.8 $\pm$ 2.0 \\
  7 & EED &             &  26 &  13.9 &  286.2 &  65.3 &  200.0 & -18.9 &  183.9 &  78.0 &   6.26 &  -8.18 & -10.81 & -6.10 &    0.26 &         17.1 $\pm$ 4.0 \\
\enddata
\tablenotetext{a}{Half-mass radius, computed in three dimensions. This is different from the use of a 2-D on-sky r$_{hm}$ measurement  for the core of EE Dra in Section \ref{sec:res-EEDra}. This choice is made to account for the occasionally non-spherical nature of CFN subgroups, and for the fact unless investigating sub-pc scales like in central EE Dra, in most of CFN uncertainties along the radial direction have a limited impact on dispersion.}
\vspace*{0.1in}
\end{deluxetable*}

SPYGLASS-I provided a subclustering analysis for CFN, although the somewhat conservative clustering parameters chosen, combined with significant projection effects along the transverse velocity axes, limited the sensitivity. That analysis identified only two subgroups within CFN: CFN-1, which covers the entire near side of the association, and CFN-2, which is smaller, more distant, and centered around the $\beta$ Cephei system. Recent work by \citet{Szilagyi21} proposes the presence of possible substructure within CFN-1, a possibility supported by a visual inspection of our populations. 

To deepen our analysis of CFN's substructure and minimize the influence of projection effects, we perform a new clustering analysis in 5 dimensions using galactic XYZ coordinates and the $l$/$b$ transverse velocity anomaly ($\Delta$v$_{T}$), which we define as the transverse velocity minus the projected velocity vector of the cluster centre at the location of each star. After the photometric restriction of the sample to 549 candidate members (\S\ref{sec:sss}), we find a new median velocity vector for the association at \edit1{(U,V,W)=(-10.13, -12.10, -5.09)} km s$^{-1}$ and use it to compute transverse velocity anomaly. The use of velocity anomaly minimizes projection effects while maximizing the number of stars available for clustering. Clustering directly on the UVW velocity is preferable for completely eliminating projection effects, but that choice also results in losing the roughly 70\% of our sample with no radial velocities, so the use of transverse velocity anomaly offers a compromise. 

We also apply a scaling factor of $c = 6$ pc/km s$^{-1}$ to the velocity axes to make them similar in scale to the spatial axes, following the choice made to enable clustering analyses in SPYGLASS-I. This choice was based on the typical ratios of velocity dispersion to size for groups in that work, as well as the size-velocity relations predicted by Larson's Law assuming motions of the parent cloud are preserved in the resulting stars. While the subgroups in CFN are smaller than the regions used to set that scaling factor in SPYGLASS-I, perhaps motivating the use of a smaller scaling factor, we see visibly larger scale ratios in CFN between spatial scales and velocity, motivating the use of the larger scaling factor.

Like in SPYGLASS-I, we use HDBSCAN for clustering \citep{McInnes2017}, and we apply the algorithm to the maximally broad sample of 549 candidates member stars to ensure all credible candidates can contribute to overdensities.
We use excess of mass (EOM) clustering with min$\_$samples and min$\_$cluster$\_$size both set to 7, allowing slightly more subtle clustering than performed in SPYGLASS-I. EOM clustering produces the most persistent clusters in HDBSCAN's clustering tree, i.e. clusters that emerge over the widest range of scales. In SPYGLASS-I we also included leaf clustering to define subgroups, which provides the clumps that emerge at the smallest scales on the HDBSCAN clustering tree. In the case of CFN, both clustering options produce the same clusters, so we conclude that only EOM clustering is needed to provide a comprehensive view of CFN substructure. This clustering approach identifies 7 subgroups in CFN. 
We experimented with a few other algorithms, including $k$-means, Fuzzy c-means, Spectral Clustering, and Gaussian Mixture, however we found that the clusters that result do not always reflect the structures that are visually identifiable. All of these algorithms locate clusters of similar sizes and densities, as these scales are either direct inputs into these algorithms, or arise as a consequence of another parameter like the number of clusters in $k$-means. The result is that the scales of the resulting groups are effectively defined by the user, and that groups smaller than the chosen scale tend to be merged at the same time that larger groups are fragmented, a result that definitionally excludes the sort of multi-scale structure that appears to be present in these regions. 

\begin{figure*}
\centering
\includegraphics[width=17cm]{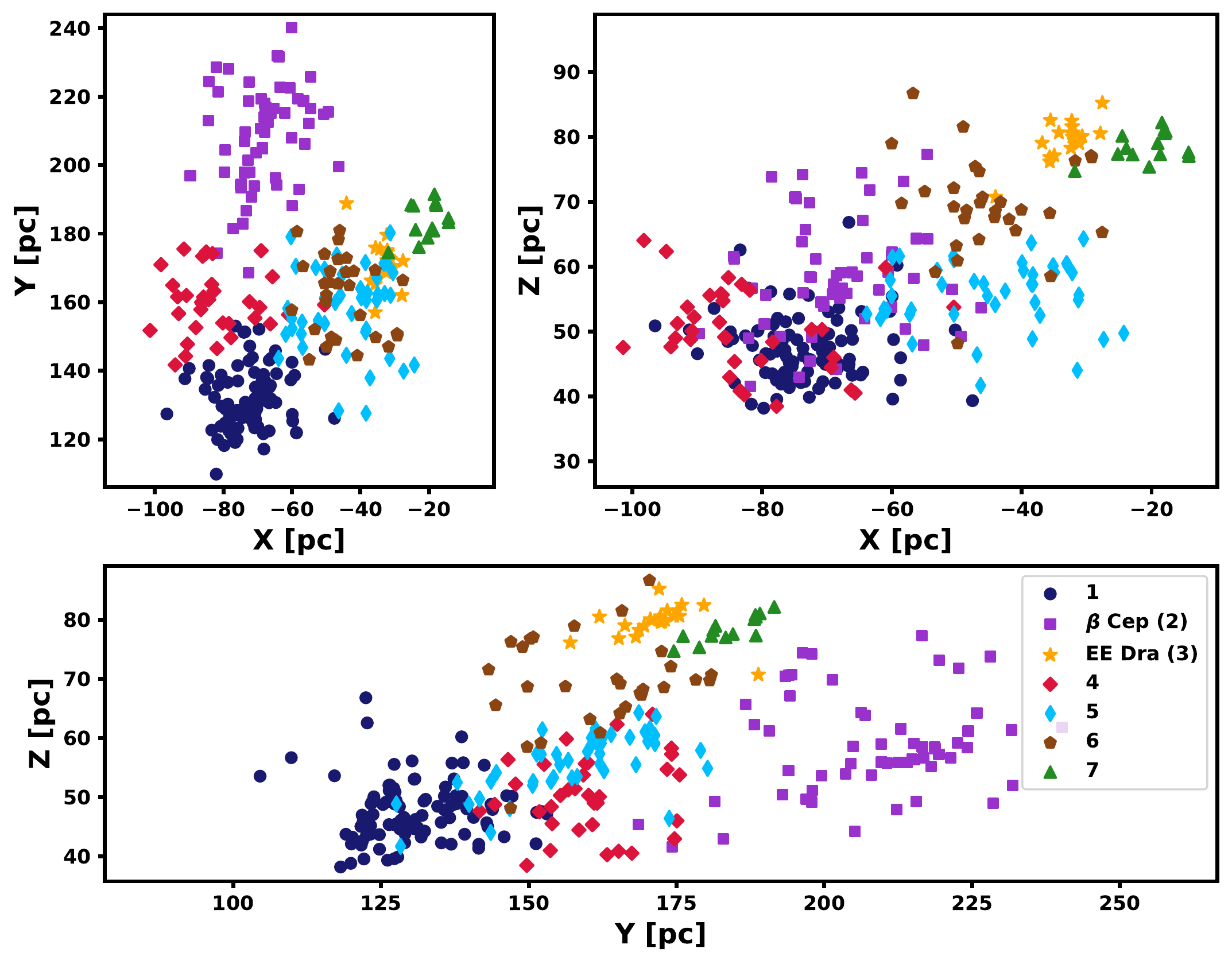}\hfill
\caption{Distribution of CFN candidate members in spatial coordinates, plotting all three axis combinations. The symbols assigned to members of each subgroup are marked in the legend in the bottom panel. }
\label{fig:xyzclustering}
\end{figure*}

\begin{figure}
\centering
\includegraphics[width=8.5cm]{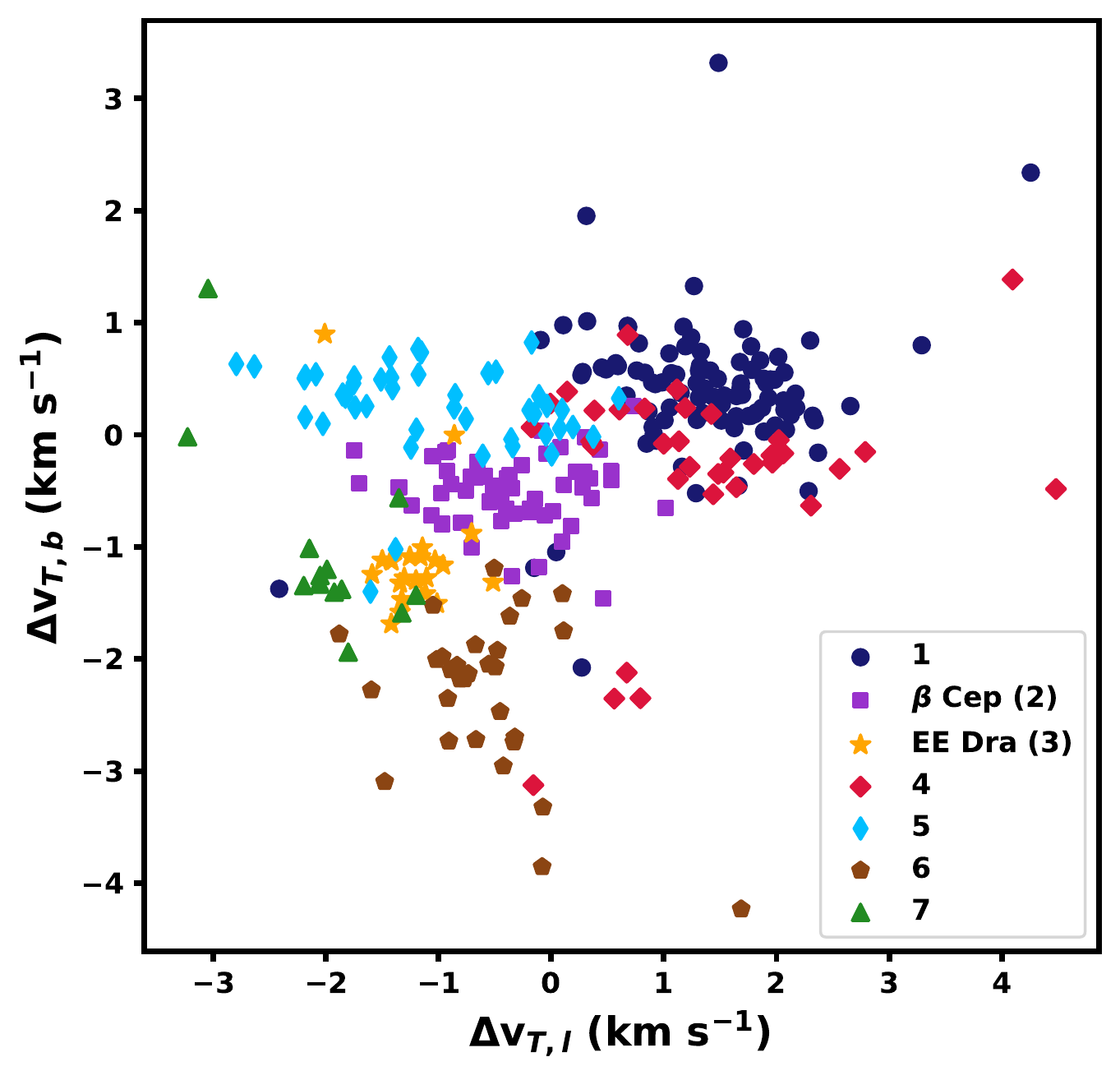}\hfill
\caption{Distribution of CFN candidate members in velocity anomaly space, marked according to their parent region. The symbols and colors are the same as those used in Figure \ref{fig:xyzclustering}}
\label{fig:velocanom}
\end{figure}

While these groups provide an excellent indication of the locations where substructure is present, HDBSCAN is designed for identifying overdensities within a noisy data set, meaning that a non-clustered background is expected. The result is that in locations where most stars are assumed to be members of a group, like in our sample, many stars are assigned to a background component despite origins in one of the identified subgroups. We must therefore reintroduce members from the HDBSCAN background into a credible parent subgroup to maintain a robust sample for analysis. We assign these outlying members to the association with the nearest core in 5-dimensional galactic XYZ/ l/b transverse velocity anomaly space. We changed the scaling factor to $c = 12$ pc/km s$^{-1}$ for this outlying member assignment, as we found that CFN subgroups often had significantly lower velocity spreads ($\sim$0.5-2 km s$^{-1}$) compared to their sizes ($\sim$5-30 pc) relative to similar objects used to define this factor in SPYGLASS-I. The use of a smaller scaling factor of $c = 6$ pc/km s$^{-1}$ in the initial cluster produced a more space-weighted clustering metric and by extension stricter conditions for inclusion in spatial coordinates compared to velocity. This produces subgroups consisting primarily of spatially central objects with a range of velocities, so the velocity-weighted outlying member assignment produced by $c = 12$ pc/km s$^{-1}$ ensures that background-assigned stars are distributed to groups with additional weight on their motions, a useful feature given the spatially ambiguous nature of these unclustered stars. 

Just over half of stars in the sample are assigned to the background by HDBSCAN's clustering, and our process of assigning outliers to the nearest subgroup neatly divides CFN between the seven clusters identified. Each subgroup appears visually distinct in space and velocity coordinates. \edit1{While not all members will have unambiguous subgroup assignment due to the occasionally overlapping intrinsic spreads of these groups in space-velocity coordinates, extreme outliers are relatively rare. Furthermore, sources with a complete 3D velocity vector are found to reliably stay together when traced back in time, enforcing common formation and suggesting that even if membership assignment is imperfect, it is unlikely to affect later dynamical analyses (see Section \ref{sec:traceback}). }These groups all occupy a tightly-distributed region in velocity space, with velocity anomaly differences between subgroup centres not exceeding 4 km s$^{-1}$. The spatial distribution of stars is more dispersed, however, as the association as a whole spans roughly 100 pc in spatial coordinates. The distributions of each subgroup's members in velocity anomaly and spatial coordinates are shown in Figures \ref{fig:xyzclustering} and \ref{fig:velocanom}. We also provide basic properties for each subgroup in Table \ref{tab:sgprops}, including the number of stars assigned to each population, their positions in both on-sky and galactic Cartesian coordinates, their half-mass radii in three dimensions (calculated using the masses from Section \ref{sec:masses}), and their galactic UVW velocities. While the resulting clusters are not necessarily definitive, as the boundaries between populations and generations are not expected to be clean, our results nonetheless provide coherent populations through which to investigate stellar properties. 

Our new subgroup CFN-2 effectively re-identifies CFN-2 as defined in SPYGLASS-I, which we hereafter refer to as the $\beta$ Cephei group ($\beta$ Cep) after the star system of the same name that it contains. The remaining six subgroups are resolved from within SPYGLASS-I's CFN-1 region. We hereafter use the CFN-1 label to refer to the subgroup occupying the core of the association of the same name defined in SPYGLASS-I, which has a similar extent to the Cepheus association as defined in \citet{Klutsch20}. The new subgroups are given new subgroup IDs. Two of these new subgroups, CFN-5 and CFN-6, lie within the HD 190833 subgroup defined by \citet{Szilagyi21}, separated from each other by $\sim$2 km s$^{-1}$ in $\Delta$v$_{T}$. The remaining clumps include CFN-4, which forms a tenuously separated extension of the CFN-1 core region, and CFN-3 and CFN-7, which are the two subgroups farthest from the galactic plane. CFN-3, centered around the star EE Draconis, is notable for the very dense configuration of its core. We hereafter refer to this group as the EE Draconis cluster (EE Dra), and investigate its status as a possible virialized open cluster in Section \ref{sec:res-EEDra}. The expansive network of subgroups we uncover demonstrates that even more substructure is present in the CFN than has been previously claimed.

\subsection{Ages}

\begin{figure}
\centering
\includegraphics[width=7.0cm]{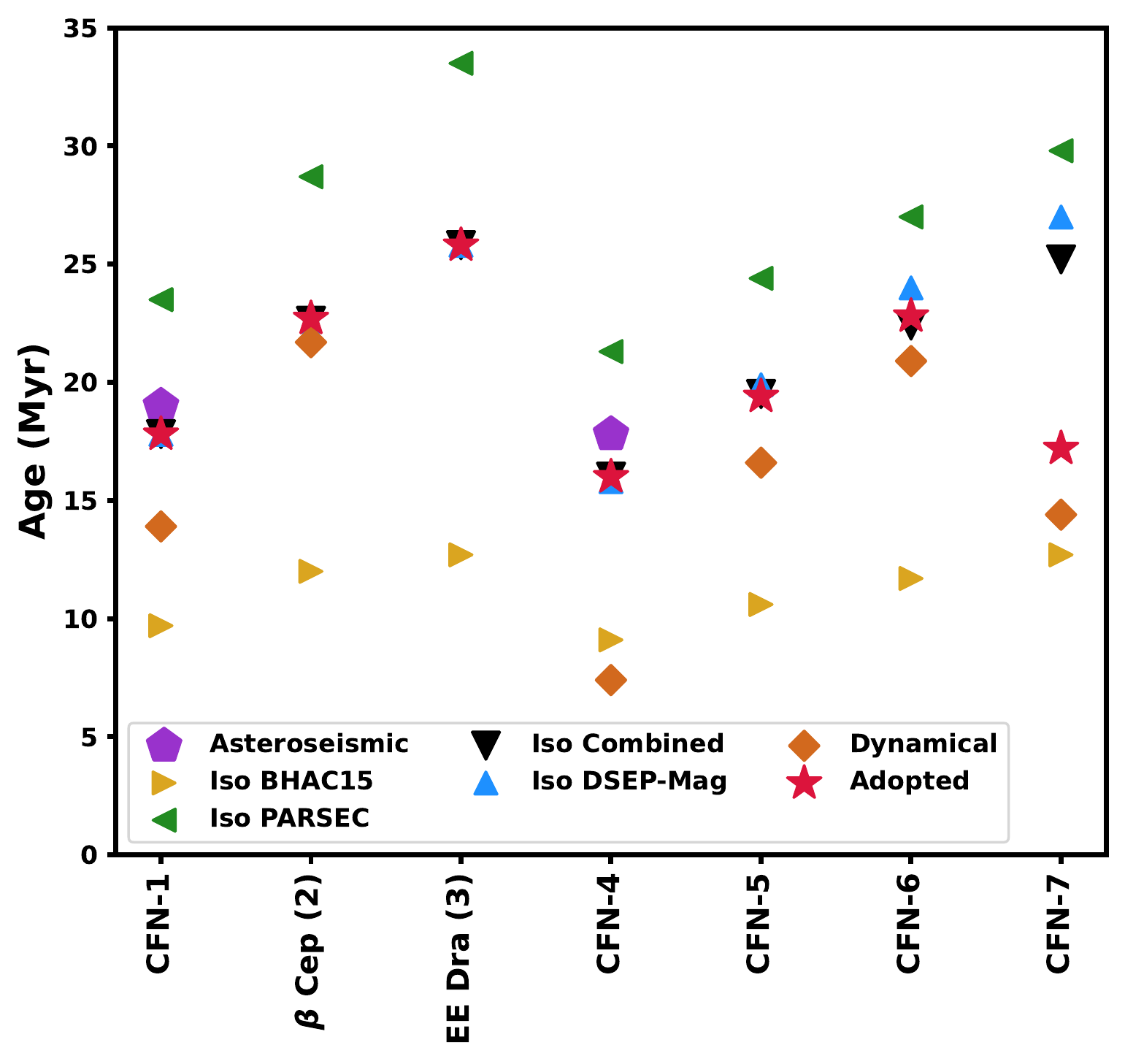}\hfill
\caption{Ages for each CFN subgroup, as calculated using different methods. The legend provides the markers for each method. These include three different isochronal ages (triangles) plus a combined result, dynamical ages (diamonds) for all groups but the virialized EE Dra clsuter (see Section \ref{sec:res-EEDra}), and Asteroseismic ages (pentagons) where available. Finally, we mark adopted ages using red stars.}
\label{fig:allages}
\end{figure}

\edit1{The age coverage of CFN to date is limited, including just a 10-20 Myr age from lithium depletion and isochrones for the CFN core reported in \citet{Klutsch20}, a 24 Myr isochronal age solution in SPYGLASS-I, and a 9 Myr isochronal age for notable CFN member $\beta$ Cep in \citet{Nieva14}. Our widespread coverage of CFN with new spectroscopic measurements, combined with the existing photometry and astrometry, opens up multiple paths for measuring both the absolute age of the association and relative ages between subgroups, greatly deepening and refining this coverage.} Through the use of spectroscopic observations, we trace lithium depletion, while we use Gaia photometry for isochronal age estimates. The combination of Gaia astrometry and our own radial velocity measurements allows for dynamical age estimates, using the dispersal of the association and its subgroups to infer the times of formation. TESS photometry reveals that several stars in CFN also pulsate as delta Scuti ($\delta$\,Sct) stars, which we model asteroseismically for an independent fourth age source. Through the combination of these diverse methods of age estimation, we provide robust age measurements, while uncovering the patterns with which stars formed and star formation progressed across the association. In Figure \ref{fig:allages}, we compile the regional age estimates gathered through various methods, including the final ages we eventually adopt.

\begin{deluxetable*}{cccccccc}
\tablecolumns{7}
\tablewidth{0pt}
\tabletypesize{\scriptsize}
\tablecaption{Age solutions for the seven subgroups identified in CFN, alongside the combined ages using all three isochronal models, and the final adopted combined ages. Asteroseismic ages are only available for 3 CFN members, so we only report these ages where robustly-assigned member stars have TESS photometry. All ages are in Myr.}
\label{tab:ages}
\tablehead{
\colhead{Group} &
\colhead{Dynamical} &
\colhead{Asteroseismic} &
\multicolumn{4}{c}{Isochronal} &
\colhead{Adopted} \\
\colhead{} &
\colhead{} &
\colhead{} &
\colhead{PARSEC} &
\colhead{BHAC15} &
\colhead{DSEP-Magnetic} &
\colhead{Combined} &
\colhead{}
}
\startdata
1 &     13.9 $\pm$ 3.5 &     19.0 $\pm$ 1.5 &    23.5 $\pm$ 1.2 &     9.7 $\pm$ 0.4 &    17.8 $\pm$ 0.9 &     17.9 $\pm$ 0.8 &         17.8 $\pm$ 0.8 \\
2 &     21.7 $\pm$ 5.6 &                    &    28.7 $\pm$ 2.0 &    12.0 $\pm$ 0.7 &    22.2 $\pm$ 1.5 &     22.6 $\pm$ 1.3 &         22.7 $\pm$ 1.3 \\
3 & Undef \tablenotemark{a} &               &    33.5 $\pm$ 4.2 &    12.7 $\pm$ 1.2 &    25.8 $\pm$ 3.6 &     25.8 $\pm$ 2.7 &         25.8 $\pm$ 2.7 \\
4 &      7.4 $\pm$ 1.2\tablenotemark{b} &     17.8 $\pm$ 2.2 &    21.3 $\pm$ 3.3 &     9.1 $\pm$ 1.0 &    15.8 $\pm$ 1.7 &     16.0 $\pm$ 1.5 &         16.0 $\pm$ 1.5 \\
5 &     16.6 $\pm$ 2.3 &                    &    24.4 $\pm$ 1.8 &    10.6 $\pm$ 0.9 &    19.9 $\pm$ 2.0 &     19.5 $\pm$ 1.4 &         19.4 $\pm$ 1.3 \\
6 &     20.9 $\pm$ 3.6 &                    &    27.0 $\pm$ 2.5 &    11.7 $\pm$ 1.3 &    24.0 $\pm$ 3.6 &     22.4 $\pm$ 2.3 &         22.8 $\pm$ 2.0 \\
7 &     14.4 $\pm$ 3.6 &                    &    29.8 $\pm$ 4.3\tablenotemark{c} &    12.7 $\pm$ 1.7\tablenotemark{c} &    27.0 $\pm$ 5.2\tablenotemark{c} &     25.2 $\pm$ 3.4 &         17.1 $\pm$ 4.0 \\
\enddata
\tablenotetext{a}{Subgroup 3 is a gravitationally bound open cluster, and therefore cannot produce a reliable dynamical age}
\tablenotetext{b}{Solution likely not reliable: uses only 3 stars, and differs significantly from formation sequence from isochrones}
\tablenotetext{c}{Solution weighed upon heavily by older outliers, younger age probable}
\vspace*{0.1in}
\end{deluxetable*}

\subsubsection{Dynamical Age} \label{sec:dynmage}

In gravitationally unbound associations, stars disperse after formation, and tracing these stars back to their tightest past distribution can reconstruct the likely time of formation. Dynamical ages have no model-dependence, so they can provide robust results in both absolute and relative terms, with the caveat that the presence of gas early in formation will delay the onset of stellar dispersal by approximately 2-4 Myr \citep[e.g.,][]{Guszejnov22}. To compute dynamical ages, we implement a simple traceback routine and use the past configurations of association members to compute dynamical ages. Our traceback sample includes all stars that have non-Gaia radial velocities with sub-km s$^{-1}$ uncertainties, excluding sources with any evidence for a companion, resolved or unresolved. We use galpy for our traceback, using the module's numerical integration routine with the \texttt{MWPotential2014} Milky Way potential model \citep{Bovy15}. This result shows clear convergence of the association to a tighter stellar distribution well before the present day, suggesting that the association, at least as a whole, is indeed dispersing from the time of formation. However, the past convergence of CFN members does not appear to all occur at the same time, suggesting the need for ages at the subgroup level. We therefore only provide dynamical ages for subgroups, as the complex substructure of the region casts doubt on the reliability of a single-age solution. 

\begin{figure}
\centering
\includegraphics[width=7.5cm]{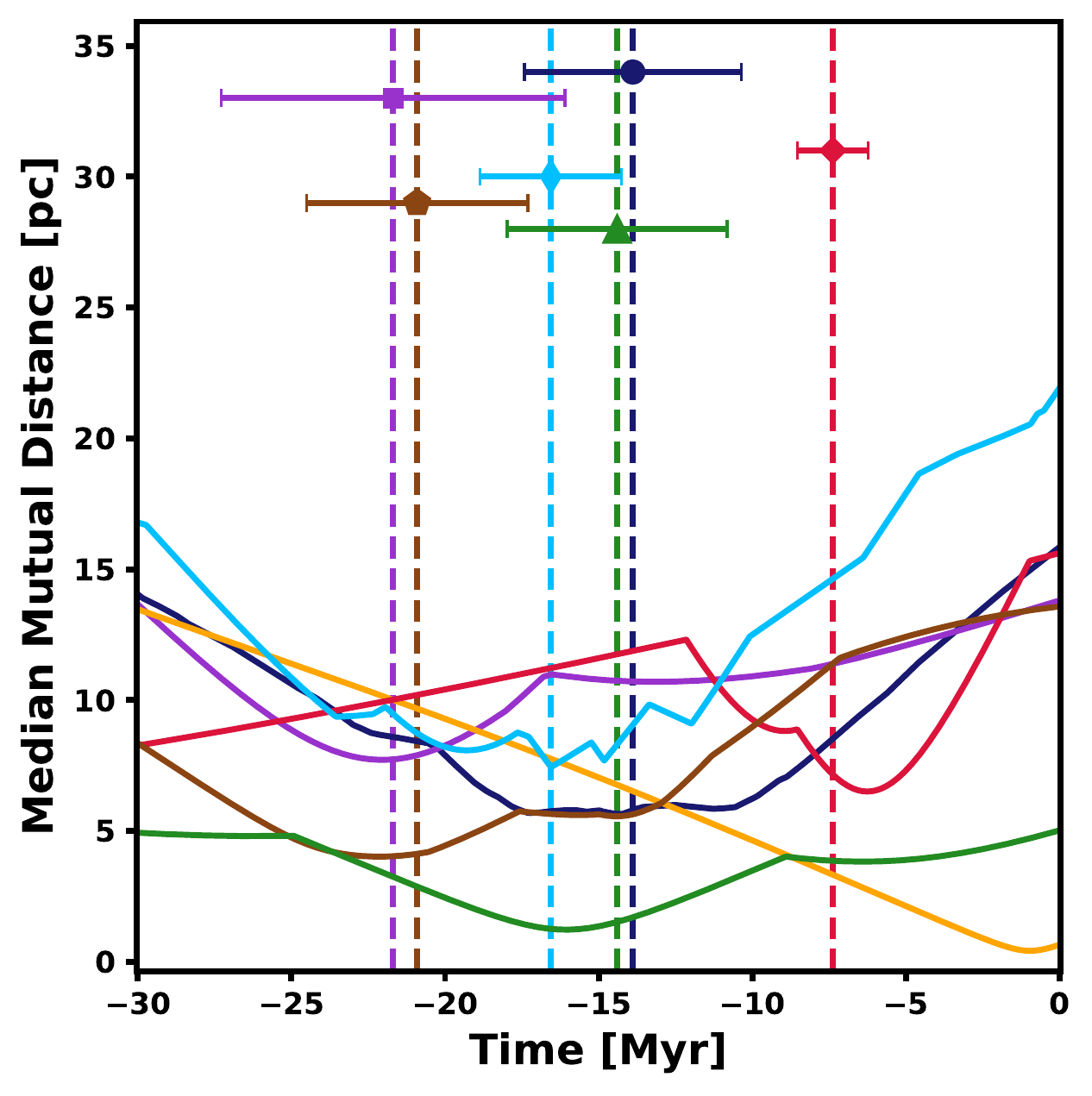}\hfill
\caption{An illustration of the dynamical ages in CFN. The curves show the median of the median distances from each star to other members, which can be seen as a proxy for the extent of each group as a function of time. The vertical dotted lines and markers indicate the dynamical age solution for each group. The icons and colors used to represent the subgroups match those in Figure \ref{fig:xyzclustering}. Note that CFN-3 (EE Dra) is a bound cluster that does not converge at an earlier date during traceback, making a dynamical age impossible.}
\label{fig:dynmages}
\end{figure}

Our age computation method applied to each subgroup is based on the mutual distances between members at each time step in our traceback. For each time step and star, we compute the median distance to all other members of the subgroup, and then for each star in the subgroup, we take the age with the minimum median distance to other members. Stars with minimum distance ages under 5 Myr are removed, as this age is significantly younger than the age of the association computed using other methods. These stars are interpreted as not dispersing like other members of the subgroup, but rather interloping either through ejection from another subgroup or through the errant velocity contributed by an unseen companion. We then compute 3-sigma clipped median and standard deviation values from the list of minimum distance ages within each subgroup, which are interpreted as the age and uncertainty of the subgroup. The results are provided in Table \ref{tab:ages}. 

In Figure \ref{fig:dynmages} we plot the dynamical age solutions and their uncertainties for each subgroup. Alongside these solutions, we provide a curve showing the running subgroup-median of each star's median distance to other subgroup members as a function of time (median mutual distance). The median mutual distance curve can be interpreted as a measure of the subgroup's extent over time, which should have a minimum around the time of formation. As expected, the dynamical age solutions align closely with the minima of those median mutual distance curves. We also include the dynamical ages in Figure \ref{fig:allages} for comparison to ages computed using other methods. 

One subgroup, EE Dra, appears to be at its most compact configuration now, as traceback only finds a larger cluster size in the past. As mentioned in Section \ref{sec:substructure}, EE Dra is much more compact than the other groups, with a core region less than one degree across in the plane of the sky. This suggests that this group is actually a virialized open cluster, making the dynamical age undefined. We investigate the status of EE Dra as an open cluster in Section \ref{sec:completeness} through a simple virial analysis of the group. Two other subgroups, CFN-4 and CFN-7, have weaker dynamical ages due to their sample size, both being based on only three stars. \edit1{CFN-4 also has a median mutual distance curve in Figure \ref{fig:dynmages} which only has a downward slope towards the solution in the region immediately surrounding the age solution, suggesting that this age in particular may lack coherence.} As such, we condition the use of \edit1{these ages} on agreement with results using other methods, as discussed in Section \ref{sec:agesum}. 

\subsubsection{Lithium Depletion}

\begin{figure*}
\centering
\includegraphics[width=9.0cm]{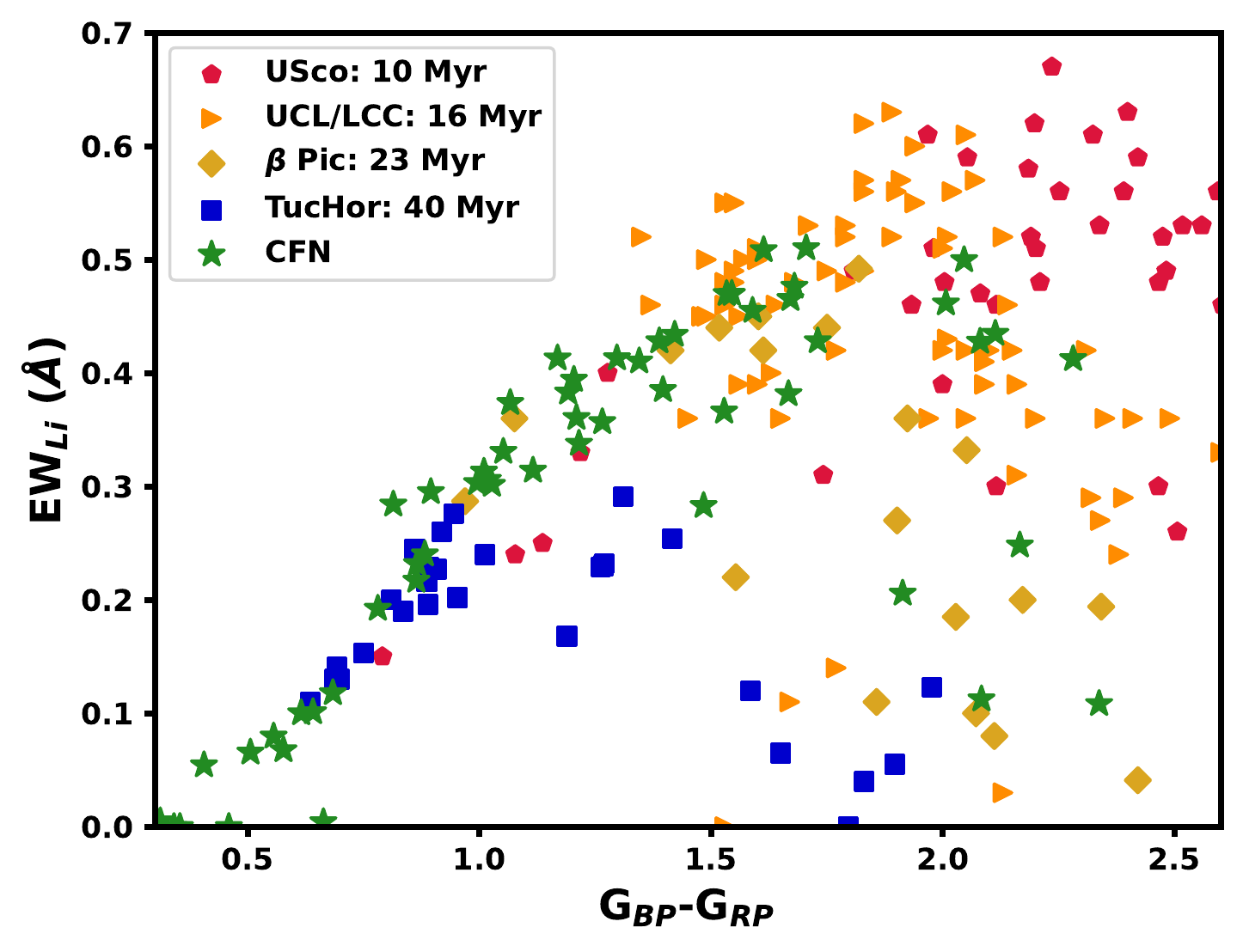}\hfill
\includegraphics[width=9.0cm]{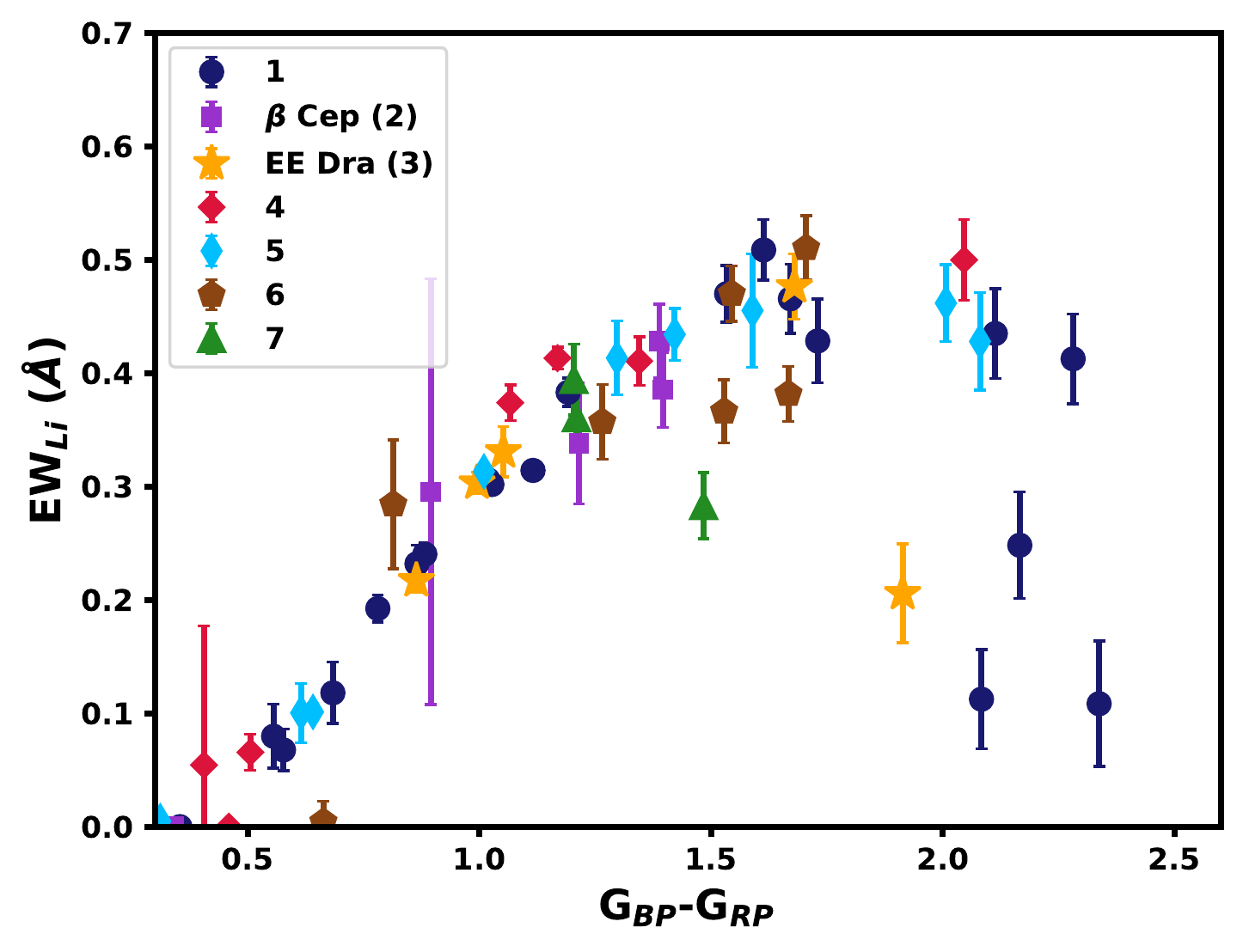}\hfill
\caption{Plots illustrating the lithium depletion sequence in CFN. In the left panel we show CFN's members as green stars, plotted against four other associations, colored by their age: Tuc-Hor, shown as squares, $\beta$ Pic, shown as diamonds, UCL/LCC, shown as triangles, and Upper Sco, shown as pentagons. The right panel focuses on only the CFN members, showing the Li EW of each star, colored by its subgroup membership. The icons and colors used to represent the subgroups match those in Figure \ref{fig:xyzclustering}}
\label{fig:Li}
\end{figure*}

Our spectroscopic sample stops just short of the lithium depletion boundary (LDB), which is often considered to provide the gold standard for age dating due to the consistency of its behavior across various models \citep[][]{Binks14}. We do however have extensive coverage of the lithium sequence for higher-mass objects, especially in the K to early M regime where lithium depletes more slowly through deep convection into the inner regions of a star where it can fuse \citep{Bodenheimer65}. Despite the more gradual depletion making the change to the lithium sequence as a function of age more subtle, these more massive stars can nonetheless provide strong age constraints through comparison to other firmly dated associations. 

We present the lithium depletion sequence for CFN in figure \ref{fig:Li}. We also plot other associations with similar ages: Tucana-Horologium (age 40 Myr; \citealt{Kraus14}), Beta Pictoris (age 23 Myr; \citealt{Mamajek14}), and UCL/LCC (ages 17 and 16 Myr; \citealt{Mamajek02, BAFFLES20}), and Upper Sco (age 10 Myr; \citealt{Pecaut16,Sullivan21}). Our corresponding lithium depletion data sets are drawn from \citet{daSilva09} for Tuc-Hor, \citet{Shkolnik17} for Beta Pic, \citet{Zerjal21} for UCL and LCC, and \citet{Rizzuto15} for Upper Sco. We supplement these sources with photometry from Gaia EDR3 \citep{Riello21}, which provides G$_{BP}$-G$_{RP}$ colors which can be used as a proxy for temperature or spectral type. 

The UCL/LCC data set from \citet{Zerjal21} in particular has a significant spread in Li EW as well as notable apparent contamination, and we therefore employ two cuts to minimize these issues. First, we remove all stars not located in a UCL or LCC-associated subgroup in SPYGLASS-I, restricting the sample to the highest probability members. We also remove any objects flagged as spectroscopic binaries, as companions introduce additional background light, often diluting the strength of the Li line. In other regions, less contamination is present so we do not impose any additional membership cuts, however we still cull binaries. Only unresolved binaries are a concern for Li measurements, so we impose the RUWE$< 1.2$ cut described in Section \ref{sec:binaries} to all external samples, including UCL/LCC. Our corresponding sample of CFN Li measurements follows all the membership and quality cuts imposed in Section \ref{sec:sss}, without applying our cut on unresolved binaries, which do not affect the reliability of Li EWs. 

The lithium depletion sequence for CFN closely follows the distribution for UCL and LCC, with a few exceptions that have lower lithium abundances compared to the rest of the members. The primary lithium sequence that aligns closely with LCC is populated mainly by members of the young core group CFN-1 and adjacent populations CFN-4 and CFN-5. However, some stars sit firmly below this main sequence of lithium abundances, with Li equivalent widths ranging from broad agreement with $\beta$ Pic, to abundances slightly lower than the $\beta$ Pic main sequence of Li abundance. Members of CFN-6 and CFN-7, both of which are presumed older populations, are overrepresented in these members with lower Li abundances. Lower Li abundances would provide an independent verification of their older ages, although given the small sample involved it is difficult to confirm that this is the result of older ages and not some caused by other influences, such as diluting flux from an unseen companion that bypassed our RUWE cuts or rotational contribution changing the rate of Li burning \citep[e.g.,][]{Messina16}. 

Due to the small sample size of stars in the necessary temperature range, our ability to robustly calculate subgroup-level Lithium ages is limited. However, the better-defined and better-populated young edge of the sequence which aligns with UCL and LCC does provide a fairly strong constraint on the typical ages across CFN-1, CFN-4, and CFN-5 at roughly 16-17 Myr \citep{Mamajek02}\edit1{, consistent with the 10-20 Myr age from \citet{Klutsch20} computed for a region mostly restricted to CFN-1 and using similar methods.} Furthermore, the overall spread in CFN enables a rough approximation for the overall duration of the star formation event in CFN. With the old edge of the CFN sequence lying just below the centre of the $\beta$ Pic sequence \citet{Mamajek14}, this appears to suggest an age range spanning from 16-24 Myr. This result is in broad agreement with the dynamical age estimates which, excluding the weakly-defined CFN-4 dynamical age, also show a similar spread of just under 8 Myr. While the absolute age is somewhat older from our lithium depletion age compared to the dynamical ages, that can be explained by a delay in dispersal caused by the presence of gas immediately following formation. We discuss the consequences of this explanation in Section \ref{sec:agesum}.  

We cross-checked our age solution for the association using the {\tt BAFFLES} package \citep{BAFFLES20}, which uses empirical measurements to compute age probability distributions based on measurements of B-V color and lithium equivalent widths. Like for collecting radial velocities, we collected B-V colors using Simbad and Vizier queries. {\tt BAFFLES} found an age for the population centered around 10 Myr. While this is notably younger than the lithium depletion age estimates reached by inspection, it does agree with the validation age that the {\tt BAFFLES} team reaches for UCL, enforcing the similarity between most of CFN's lithium depletion sequence and that of UCL and LCC.

\subsubsection{Isochronal Ages} \label{sec:isoages}

\begin{figure*}
\centering
\includegraphics[width=17.0cm]{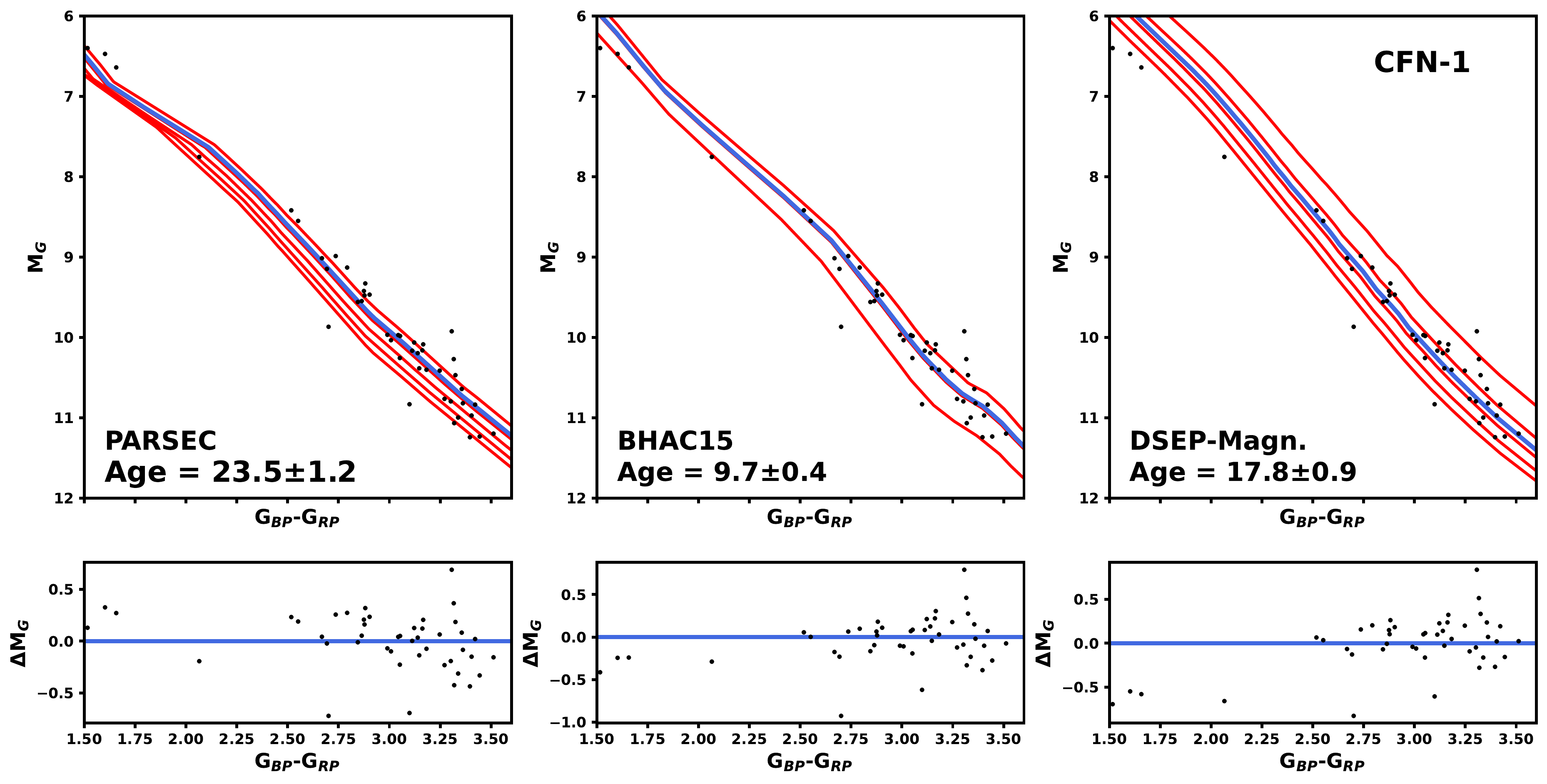}\hfill
\caption{Age fits for the CFN subgroups. The top row provides a CMD of the fit members of the subgroup. Stars included in the fitting are shown as black dots, and the best fit result is in blue. We also provide a range of isochrones for comparison, which from top to bottom are: 20, 25, 30, 35, and 40 Myr for PARSEC, 8, 10, and 15 Myr for BHAC15, and 10, 15, 20, 25, and 30 for DSEP-Magnetic. The bottom row provides residuals between the best fit model and photometry. CFN-1 is provided as an example, the full figure set is provided in the online-only version.}
\label{fig:isoagefits}
\end{figure*}

Isochronal age estimates are perhaps the most accessible age calculation method available, requiring only distances and magnitudes, both of which are readily available through Gaia. This method is especially useful for computing relative ages and age sequences, as slight but systematic differences in the height of the pre-main sequence above the main sequence can provide robust indications of when one population is older or younger than another. However, especially on the pre-main sequence, dramatic differences in age solutions can exist between models, making the systematic uncertainties of the absolute age quite large. To provide a robust view that properly captures possible variations between models, we gathered three different models: PARSEC v1.2S \citep{PARSECChen15}, BHAC15 \citep{BHAC15}, and DSEP-magnetic \citep{Feiden16}. All isochrones used in this fitting assume solar metallicity, which is a fairly consistent feature of nearby young associations \citep[e.g.][]{Almeida09}. 
Due in part to the ubiquity of assuming solar metallicity in the isochrones of young populations \citep[e.g.,][]{Herczeg15}, the options for different metallicity choices are limited, so focusing on solar metallicity results allows us to explore a wider range of models.

For each CFN subgroup, we compute a best fit age according to each model. We fit all isochrones to the corresponding stellar population using the routine described in Section 3.5 of SPYGLASS-I, which used least-squares optimization with age as the fit parameter on an isochrone grid spanning the pre-main sequence with 1.2$<$G$_{BP}$-G$_{RP}<$4. DSEP-magnetic isochrones did not quite reach down to the red limit of this color range, so we limited that fit to G$_{BP}$-G$_{RP}<$3.6. All photometry was gathered from Gaia EDR3 \citep{Riello21}, converted into an absolute magnitude $M_G$ using the Gaia parallax \citep{EDR3Astro_Lindegren21}, and dereddened using \citet{Lallement19} reddening estimates. We also removed stars with RUWE$>$1.1, which restricts the fitting to a region of parameter space in which \citet{Bryson20} finds a negligible contribution for binaries. 

We plot the isochronal solution for each subgroup using all three models in Figure \ref{fig:isoagefits}. There we show the best fit solution, the members included in the fit, and set of different-age isochrones for comparison. We also plot residuals between the stellar photometry and models in the bottom row of the figure, which provides a view of the dispersion in stellar properties. The resulting ages are compiled in Table~\ref{tab:ages}, and they are also included in Figure \ref{fig:allages}, where they can be compared to our dynamical age solutions. 

We first note the wide range of age results that isochrone fitting produces. The BHAC15 model fits span \edit1{$\sim$9-13 Myr}, a range over three times smaller than that of the PARSEC isochrones, which span \edit1{$\sim$21-34} Myr. It is no coincidence that older age solutions correlate with a wider age spread, as stellar evolution is more gradual at older ages, making the same amount of displacement on a CMD infer a larger difference in age. With age ranges of $\sim$14-22 Myr implied by dynamical ages, and $\sim$16-24 Myr implied by lithium depletion, it is clear that the closest match is provided by the DSEP-magnetic isochrones, which range between $\tau \sim$ 16-27 Myr. This result still provides a wider range of ages than what our lithium and dynamical ages predict, but given the systematic uncertainties it is consistent. Further discussion of the differences between age solutions based on different isochrone models can be found in \citet{Herczeg15}, while \citet{Pecaut16} and \citet{Feiden16} provide additional comparisons between isochrone models and lithium depletion ages. 

\subsubsection{Asteroseismology}

\begin{figure}
\centering
\includegraphics[width=0.97\columnwidth]{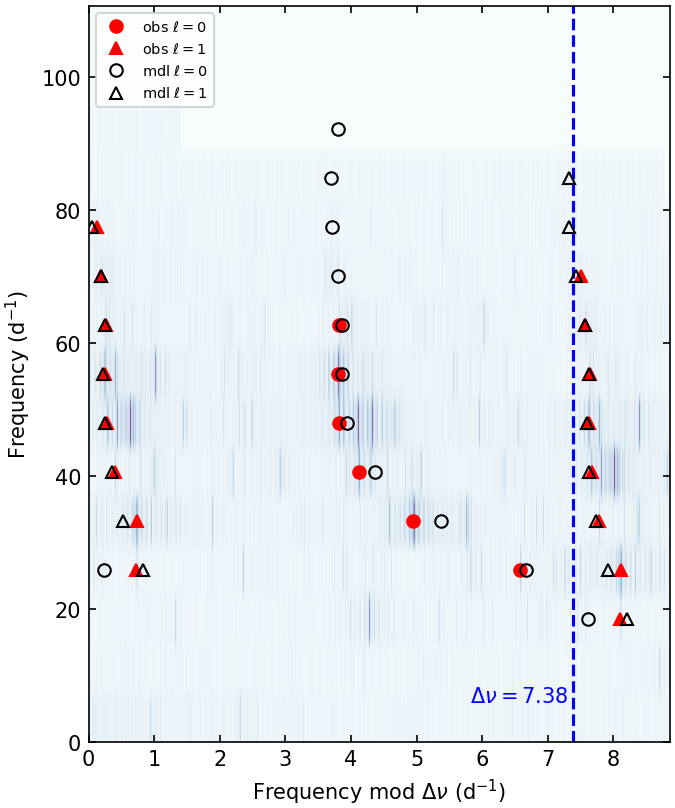}\hfill
\caption{Échelle diagram with model frequencies at 14\,Myr for TIC\,376872090. Radial modes are shown as circles, dipole modes are triangles. Unidentified peaks correspond to rotationally split modes or higher degree modes that were not modelled. The right-hand side of the diagram (with $x>\Delta\nu$) is repeated for clarity.}
\label{fig:ech}
\end{figure}

The pulsations of young $\delta$\,Sct stars sometimes show regular patterns that allow modes to be readily identified \citep{beddingetal2020}. When modelled, those modes offer asteroseismic ages with a precision as fine as 7\% in some cases \citep{murphyetal2021a}. We found three stars in CFN whose TESS data\footnote{\edit1{TESS Observations for these stars are available through the Mikulski Archive for Space Telescopes (MAST) at the Space Telescope Science Institute, and can be accessed via \dataset[10.17909/t9-nmc8-f686]{https://doi.org/10.17909/t9-nmc8-f686}}} reveal these regular patterns and we modelled them in this work, assuming a solar metallicity (initial metal mass fraction $Z_{\rm in} = 0.014\pm0.015$; \citealt{asplundetal2009}). We identified the modes similarly to \citet{murphyetal2021a,murphyetal2022a}:  that is, for each star we initially searched for a value of the asteroseismic large spacing ($\Delta\nu$) that creates vertical ridges in the star's \'echelle diagram. We then identified modes along a strong curved ridge up the centre of the \'echelle diagrams as radial modes, and in two of the three stars we also identified a vertical and almost straight dipole ridge. Since we model individual mode frequencies, the results are insensitive to the particular initial value of $\Delta\nu$. An example mode identification is shown for TIC\,373018187 (Gaia EDR3 2233485145423628928) in Fig.\,\ref{fig:ech}.

We computed stellar evolutionary models with MESA (r15140; \citealt{paxtonetal2011,paxtonetal2013,paxtonetal2015,paxtonetal2018,paxtonetal2019}) and stellar pulsation calculations for $\ell=0$ and 1 modes with GYRE (v6.0.1; \citealt{townsend&teitler2013}). As in \citet{murphyetal2022a}, we applied a helium enrichment ratio of dY/dZ = 1.4 with $Y_0=0.2800$, which at the applied metallicities results in an initial helium mass fraction of 0.2776--0.2818. 
\edit1{
For each star independently, we restricted models to a 3$\sigma$ classical error box (i.e. $T_{\rm eff}$ and $\log L$) around the parameters in Table~\ref{tab:seismo}. The ranges of $T_{\rm eff}$ are based on the spread in values across literature sources \citep{Ammons06,Anders19,Bai19}, with a floor of 250 K to better represent typical Teff uncertainties for A stars \citep{Niemczura15}. Luminosity values are based on $m_V$ from \citet{Hog00}, with bolometric corrections from \citet{Flower96}, distance corrections using Gaia EDR3 parallaxes, and \citet{Lallement19} reddening corrections. 
}
Evolutionary tracks were calculated for a mass range of \edit1{1.35--1.95\,M$_{\odot}$ at steps of 0.01}\,M$_{\odot}$, while metal mass fractions were sampled in steps of 0.0005 between $Z_{\rm in} = 0.0125$ and 0.0155. All three stars had $\chi^2$ minima within this grid, but we note that deeper global minima might exist at different metallicity. The evolutionary tracks were sampled very finely in age (steps of \edit1{$1.25\times10^4$\,yr, with outputs every 4} steps), and linear interpolation was used to ensure that consecutive models had changes in $\Delta\nu$ no larger than 0.01\,d$^{-1}$. The models were all non-rotating. More details on this grid and a wider application to many stars will be presented in a forthcoming paper (Murphy et al. in prep.).

\begin{table*}
\centering
\caption{Star IDs from the TESS Input Catalogue and Gaia EDR3 for the three asteroseismic targets, along with the effective temperature and luminosity constraints used in the modelling (one sigma uncertainties given). The obtained stellar ages are given in the penultimate column with their one sigma random uncertainties, followed by subgroup membership in the final column.}
\label{tab:seismo}
\begin{tabular}{cccccl}
\toprule
TIC & Gaia EDR3 & $T_{\rm eff}$ & $L$ & age & subgroup \\
& & K & $L_{\odot}$ & Myr & \\
\midrule
373018187 & 2233485145423628928 & $7950\pm250$ & $7.030\pm0.200$ & $17.76\pm2.16$ & CFN-4 \\
376872090 & 2267621343629966336 & $7980\pm250$ & $8.333\pm0.250$ & $14.18\pm0.25$ & CFN-6\tablenotemark{a} \\
429019921 & 2283081473548728704 & $8300\pm300$ & $8.374\pm0.300$ & $19.02\pm1.46$ & CFN-1 \\
\bottomrule
\end{tabular}
\tablenotetext{a}{This star is highly outlying within CFN-6. CFN-4 is a comparably good fit, and would agree much more closely in terms of age}
\end{table*}

The best-fitting stellar age belongs to the model with the lowest $\chi^2$, calculated as the sum of squared frequency differences between observed and model mode frequencies. We normalised the $\chi^2$ values, such that the best-fitting model for each star has $\chi_n^2=1$, then calculated \edit1{the random} uncertainties as the \edit1{standard deviation} of models with $\chi_n^2<=2$, taking the median as the reported age. The age zero-point is defined as the point where the central core temperature reaches $9\times10^5$\,K. The choice of age zero-point within commonly used options has a negligible effect on reported ages (of order 0.01\,Myr; see \citealt{murphyetal2021a} for a discussion), whereas larger systematic uncertainties exist through the use of non-rotating models \citep{murphyetal2022a}. We estimate the sum of systematic uncertainties to reach $\sim$2\,Myr. This is comparable to the random uncertainties but becomes unimportant for comparing asteroseismic ages relative to each other.

The obtained ages 
\edit1{
are given in Table~\ref{tab:seismo}.
}
TIC\,373018187 and TIC\,429019921 have well-defined membership in subgroups CFN-4 and CFN-1, respectively, putting them well within the range of ages suggested so far through isochronal, dynamical, and lithium depletion ages. We therefore take the age measurements for TIC\,373018187 and TIC\,429019921 to provide representative ages for the subgroups CFN-4 and CFN-1, respectively. TIC\,376872090 has provisional membership in CFN-6, however it fails our later RV cut and has highly outlying membership within that subgroup, making its parent subgroup unclear. If TIC\,376872090's membership in CFN-6 were accurate, this age solution would be unexpectedly low, as CFN-6 is one of the dynamically and isochronally oldest subgroups in the association, consistently producing ages older than CFN-1 and CFN-4. However, this star has a position in space-velocity coordinates not unlike some outlying members of CFN-4, a group that would align much more closely with this young age. As a result of this uncertainty, we do not use this age as a representative of any subgroup. Our group asteroseismic ages from this stellar sample are recorded in Table \ref{tab:ages}.

\subsubsection{Age Synthesis} \label{sec:agesum}

\begin{figure}
\centering
\includegraphics[width=6.0cm]{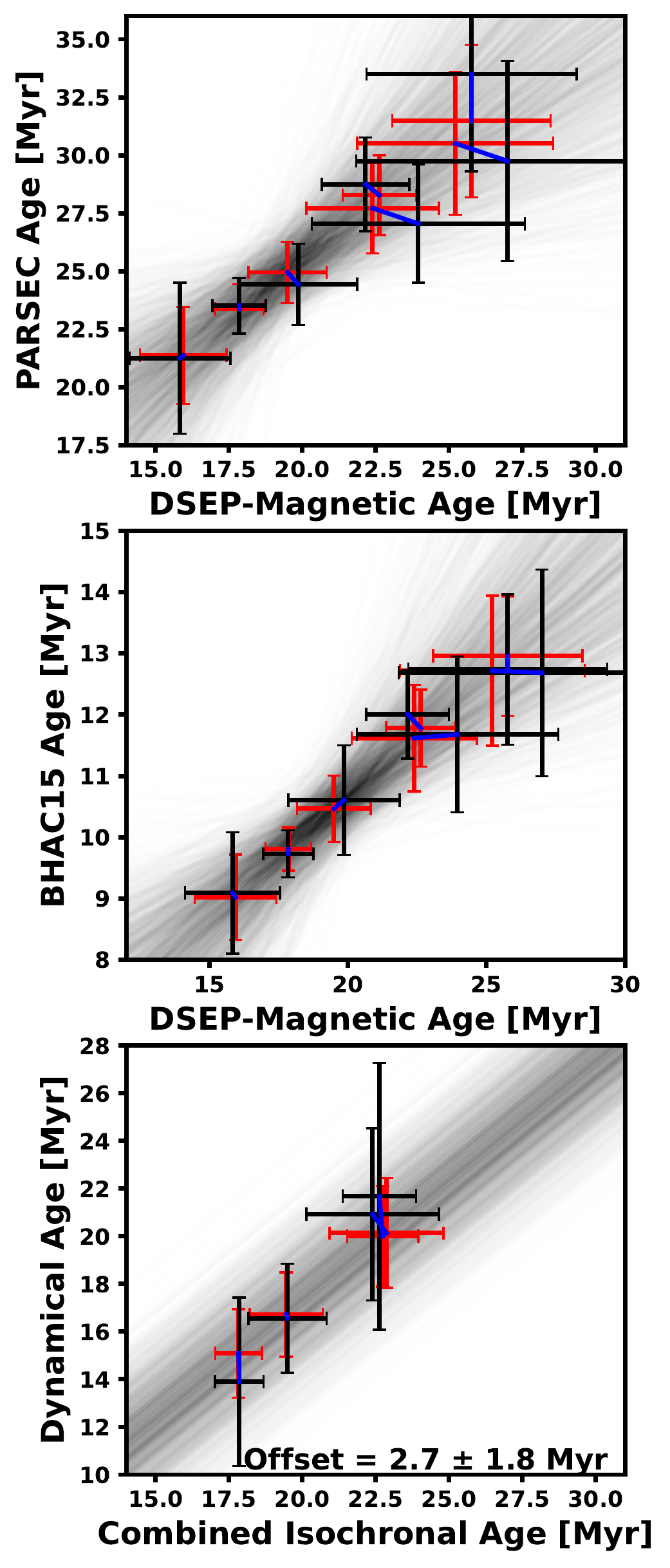}\hfill
\caption{Fits between different age solutions, which are used to produce combined ages. The input ages are shown in black, and the resulting fit ages, which follow the average line fit, are shown in red. The blue lines connect each star's input age to their refined fit age. A set of fit results from our Monte Carlo sampling is shown as the transparent dark lines, indicating the approximate range of plausible fits. The isochronal age fits, which are shown in the first two panels, are selected in three dimensions, which is why not all fits appear to connect an age set to the closest point in terms of sigma along the mean fit when projected in two dimensions. The bottom panel fits an age offset between the dynamical and isochronal ages, with the precise value included. Systematically young dynamical ages are caused by the continued presence of gas that keeps the group bound, so the offset reflects the gas dispersal timescale.}
\label{fig:unifiedages}
\end{figure}

Here we synthesize data from all age estimates used in this section, producing unified age estimates for each subgroup moving forward. We first must note the strengths and weaknesses of each age estimation method. Lithium depletion ages are generally the most widely accepted as accurate in both absolute and relative terms, however our capabilities are blunted by our small sample sizes for many subgroups, especially in the region where lithium evolves most quickly in this age range, as well as our lack of any observations at the lithium depletion boundary. This prevents robust Li ages for individual groups but provides more general age constraints for the association as a whole. Dynamical ages also provide strong model-independent age solutions, making them robust in absolute terms and especially in terms of the age range, with the caveat that they measure age from the moment of gas dispersal at which the clump becomes unbound. Isochronal ages provide robust comparative ages between different subgroups, however their model-sensitivity makes them less reliable for computing absolute ages. As such, they can be useful for setting the formation sequence but need an external anchor to make sure that the absolute ages can be trusted. Finally, asteroseismic ages are limited in their coverage as they require the presence of $\delta$ Scuti stars, which are not available in all subgroups. They are also model dependent, but are insensitive to binarity, dust (i.e. extinction and reddening), disks, peculiar motions, and to a limited extent, rotation. Asteroseismic ages therefore provide a strong independent age comparison where available. 

We first produce a combined isochronal age. All three models we employ have different shapes and have their own strengths and weaknesses in fitting the data, but they all fundamentally measure the same thing (height above the ZAMS), with model-dependent systematic offsets and scaling. We propose that by determining the offsets and scaling relationships that connect different isochronal fits, we can assess the isochronal solutions in a unified manner and refine ages. We employ an orthogonal distance regression fitting routine to incorporate uncertainties in both isochronal age axes, and fit linear relationships between the DSEP-Magnetic Ages and each of the PARSEC and BHAC15 models. The combination of these fits produces a line in 3-dimensional space, which essentially averages over the non-systematic deviations between different models. 

In Figure \ref{fig:unifiedages}, we plot these ages against one another, fit lines to these sets of points, and place the resulting fit ages along those lines. We compute uncertainties in each result by applying a Monte Carlo approach for each age solution, taking 10000 age samples drawn randomly from a normal distribution centered on the age solution and with a standard deviation equal to the uncertainty. For each of the 10000 randomized age samples, we fit a line by least-squares optimization, and locate the closest point along the line for each CFN subgroup, with the distance metric to the line normalized according to the uncertainty in the age value along each axis. We must then read off a solution from the line as a consensus age, which can be set to have scaling consistent with any of the three isochronal methods. The final age value and uncertainties are then given by the mean and standard deviation of the fits to the age sample, read off using the desired reference age scaling. DSEP-Magnetic ages clearly have the result most consistent with other methods with reliable absolute scales. These models were designed to produce agreement with lithium depletion ages, so it is not a surprise that they also agree quite closely with our lithium results. The young edge of our lithium sequence is set by 5 stars, spread across CFN-1, CFN-4, and CFN-5. The DSEP-Magnetic ages of these groups span from \edit1{15.8 to 19.9 Myr}, putting them in broad agreement with the Lithium Sequence centered around 16-17 Myr at the UCL/LCC sequence. The absolute values of the DSEP-Magnetic isochrones also agree with the asteroseismic ages for stars associated with CFN-1 and CFN-4, at 19.0 and 17.8 Myr. Due to these independent sources supporting the absolute scaling of the DSEP-magnetic isochrones, we read all solutions off from our fits according their positions along the DSEP-Magnetic Age axis. The results are provided in Table \ref{tab:ages}, and represent similar but refined versions of the DSEP-Magnetic solutions. 

Finally, we must unify this combined isochronal age with the dynamical ages. As we show in the bottom panel of Figure \ref{fig:unifiedages}, we take the same fitting approach for unifying the dynamical and isochronal ages that we used for the different dynamical ages. We also lock the slope to 1, reflecting the expectation that our corrected and unified isochronal ages should agree with the dynamical ages in terms of the duration of the star formation event. The only expected difference between the two should be caused by the gas dispersal timescale which delays the dispersal measured by dynamical ages. It is therefore a delay in absolute ages caused by the gas dispersal timescale that we are effectively fitting for. However, since not all subgroups have good age solutions in both dynamical and isochronal ages, we must first remove these problematic groups to avoid them skewing the fits.

Two groups have weak or non-existent dynamical ages: EE Dra and CFN-4. For EE Dra, we are unable to calculate a dynamical age in the first place due to its apparently virialized nature, which we discuss further in Section \ref{sec:res-EEDra}. CFN-4 does have a dynamical age, and both the dynamical and isochronal ages agree that CFN-4 is the youngest CFN subgroup. However, its dynamical age is far younger than the roughly CFN-1-aligned isochronal results would suggest, at only 7.4 Myr. The group has only three stars available for use in dynamical ages, and the result firmly disagrees with the isochronal ages, so we exclude it from fitting. 

Only one other group shows consistent disagreement between the formation order produced by isochronal and dynamical ages: CFN-7. Upon inspection, we find the inconsistent solution is likely a product of the small isochronal fit sample size in the subgroup. With the RUWE $<1.1$ cut for removing possible binaries, the group contains only 6 objects in the lower section of the pre-main sequence which shows the largest dynamic range as a function of age, including 4 along a straight sequence with little scatter, and the remaining two lying well below that. Due to CFN-7's close proximity to the EE Draconis cluster, which has the oldest age of all CFN subgroups according to the BHAC15 and PARSEC isochrone fitting, these low-luminosity outliers may be interlopers from EE Dra, or they may be field interlopers, the latter being especially possible here, as both outliers lie very close to the lower limit at which stars on the pre-main sequence are identified as young. Regardless of the explanation for the outliers, most of the stars on CFN-7's pre-main sequence have high photometric positions similar to stars in CFN-1 and CFN-5, the two groups it is placed between in dynamical age, suggesting that the young dynamical age solution is accurate. While the apparent presence of outliers in isochronal age adds doubt to the dynamical age as well, the photometric youth of CFN-7's low-scatter pre-main sequence supports an age similar to that of CFN-1, as the dynamical age suggests. The age solution must therefore be based on the dynamical age, corrected by an offset for gas dispersal. 

With CFN subgroups 1, 2, 5, and 6 appearing to show results that have a consistent sequence across the dynamical and isochronal approaches, we fit a straight line with slope 1 to these results, and show the result in the bottom panel of Figure \ref{fig:unifiedages}. The free parameter in the fitting is a vertical offset, which can be interpreted as the gas dispersal timescale for the dynamical ages, a value which should be around 2-4 Myr based on recent simulations \citep[e.g.,][]{Guszejnov22}). For the subgroups that have good isochronal and dynamical ages, we extract results in the same method used for isochronal ages, finding the closest point along the fit to each group, and then reading the result off according to the isochronal scaling. We make this choice due to the calibration of the isochronal results with the Lithium and Asteroseismic results, as well as the lower uncertainties for the isochronal ages compared to the dynamical ages. 
For EE Dra and CFN-4, which do not have reliable dynamical ages, we simply adopt the isochronal ages. Finally, for CFN-7, which has a reasonable dynamical age but a poor isochronal fit, we use the dynamical age, but adding on the offset that we fit for. As with the isochronal ages, the resulting combined ages have uncertainties that are determined through Monte Carlo sampling. 

The final adopted ages are provided in Table \ref{tab:ages} alongside the solutions achieved using other methods, and we also list these ages in Figure \ref{fig:allages}. These results are older than the dynamical ages by between 1 and 4 Myr, quite consistent with current timelines of gas dispersal \citep[e.g.,][]{Guszejnov22}. Most of these ages have uncertainty of order a few Myr, meaning that while the roughly 10 Myr history of star formation is well-resolved given our uncertainties, individual ages could change somewhat in future studies. Due to the weakly-established gas dispersal timescale for the dynamical ages and the highly model-dependent isochronal age solutions, we relied on the asteroseismic and dynamical ages for setting the absolute values from the other methods, which were highly limited in sample size. The systematic offsets are therefore particularly vulnerable to change, and they should improve with new observations. Further exploration of the Lithium sequence for stars at both the young and old limits of CFN's age sequence would therefore greatly improve these constraints, especially if the lithium results are extended down to the lithium depletion boundary, which would provide even more rigid constraints. The additional spectroscopic observations required to expand the lithium sequence would also provide new RVs, thereby further constraining our dynamical age results in the process. Improved screening of binaries would also have a positive effect on results from dynamical, isochronal, and lithium depletion ages, as all three of those ages can be affected by the presence of unresolved binaries, which are currently imperfectly removed using our RUWE cut \citep[e.g.,][]{Bryson20,Fitton22}. 

\subsection{CFN Star Formation History} \label{sec:traceback}

\begin{figure*}
\centering
\includegraphics[width=15.5cm]{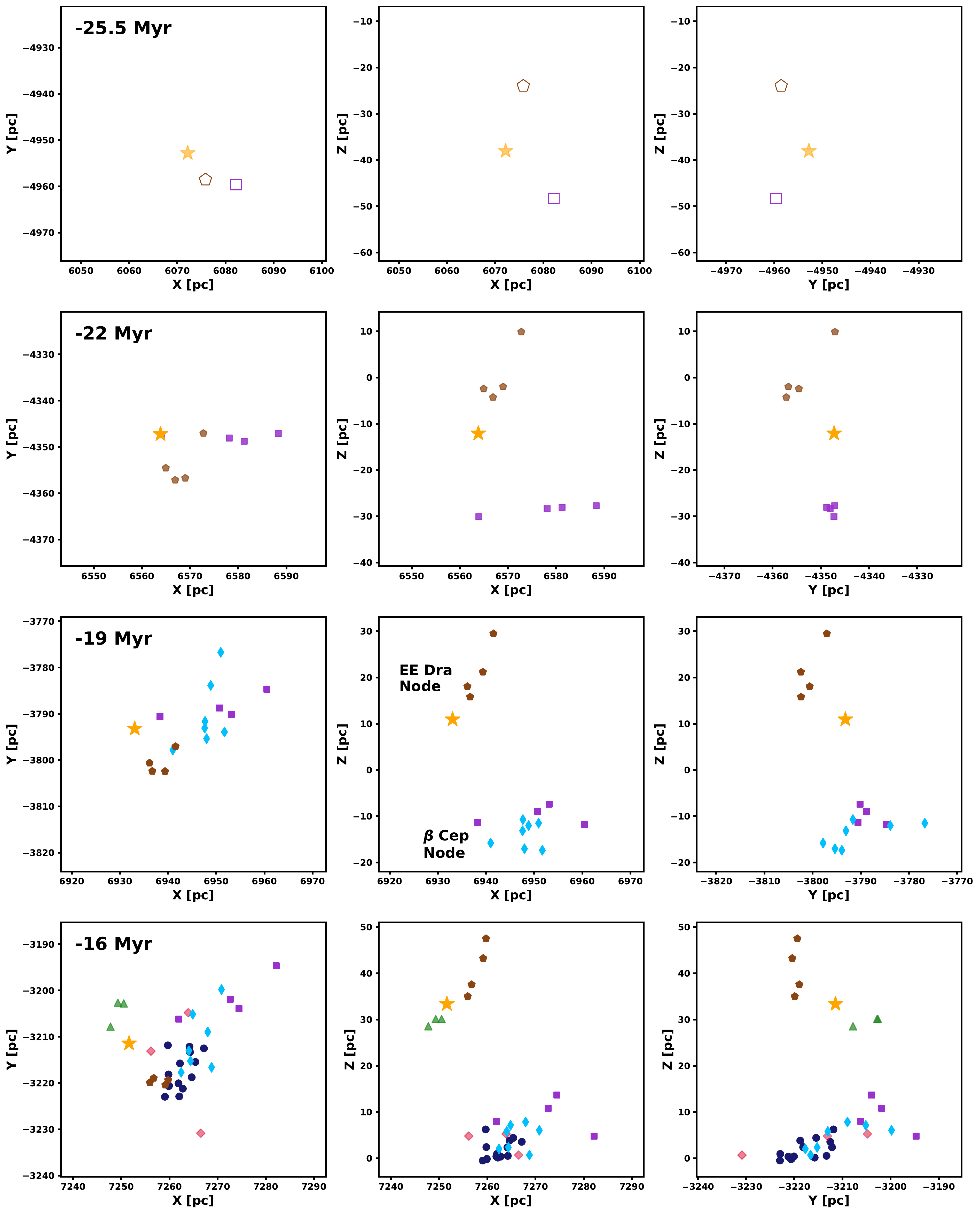}\hfill
\caption{Location of CFN members over the duration of its formation in \edit1{galactocentric} coordinates, including four time steps. The markers are the same as those used in Figure \ref{fig:xyzclustering}, with EE Dra only displayed as a co-moving centre, because as a virialized cluster its stars are not expected to converge when traced back in time (see Sec. \ref{sec:res-EEDra}). We change the transparency of the symbols from invisible to solid linearly over the age uncertainty interval of the subgroup, such that the most recently formed groups are shown as transparent. 
The first and second time steps (25.5 and 22 Myr) show the beginning and end of the star formation event forming EE Dra, $\beta$ Cep, and CFN-6, while the bottom two panels show the beginning and end of star formation in the remaining subgroups, including CFN's most populous subgroup of CFN-1. In the top panel we include the co-moving centres of $\beta$ Cep and CFN-6 to demonstrate their proximity to EE Dra immediately prior to formation. An interactive version is available in the online-only version, which includes complete traceback to the present day. \edit1{In this version, the spatial distribution of stars can be manipulated in 3D through zoom and rotation, and time can be manipulated through either the slider or the play and pause buttons which control the animation of the stars through time.}}
\label{fig:traceback}
\end{figure*}

With reliable age measurements for each subgroup synthesized through the combination of multiple age calculation methods, we can now reconstruct the entire star formation history of CFN. The combination of these ages with our dynamical traceback allows us to determine the locations of stellar populations at the moment of their formation, introducing populations to our traceback at the times they form, and showing how CFN subgroups have interacted and guided star formation over the association's history. As we will show in this section, CFN appears to have substructure in not just present day populations but also in its star formation history, with two distinct nodes that have shown continued star formation throughout the association's formation. We refer to these persistent star formation sources as the EE Draconis and $\beta$ Cephei nodes, after the groups that initiated star formation in each node.  

Figure \ref{fig:traceback} outlines the star formation sequence in Cepheus Far North which we summarize here. \edit1{The stars shown there are the same subset with high-quality kinematics used for dynamical ages, which provide the clearest view of past motions in the region and are shown in the galactocentric frame used by galpy traceback. In the top row of Figure \ref{fig:traceback}}, the star formation event begins approximately 26 Myr ago with the formation of the EE Dra cluster, at which point the cluster is within approximately 16 pc of the clouds that would soon form CFN-6 and $\beta$ Cephei. In the second row, CFN-6 and $\beta$ Cep form $\sim$3 Myr later, with EE Dra remaining close to CFN-6 while diverging slightly from $\beta$ Cep. \edit1{There is then a roughly 3 Myr age gap}, which in the third row is ended by the formation of CFN-5 at a location largely co-spatial with $\beta$ Cep. At this point, the star formation event has formed two distinctly different nodes \edit1{which are labelled in row 3 of Figure \ref{fig:traceback}}, with one containing EE Dra and CFN-6, and the other containing $\beta$ Cep and CFN-5. These nodes are well-separated from each other in the Z direction, however they overlap in X and Y. Another \edit1{2} Myr later in the the fourth row, CFN-7, CFN-1, \edit1{and} CFN-4 all form within 2 Myr of each other. CFN-1 and CFN-4 continue the star formation event in the $\beta$ Cep node with a \edit1{modest} age separation from CFN-5, producing the most populous generation in CFN, while CFN-7's formation interrupts a long period of dormancy at the EE Dra node, marking approximately \edit1{6} Myr since the previous formation of CFN-6. \edit1{An interactive version of Figure \ref{fig:traceback} is available on the online-only version of this paper, showing the full traceback to the present day.} 

A few key elements can be drawn from this discussion. First, we notice that especially later in the star-forming event, stars form in two distinct nodes, which appear to continue to form stars long after the two nodes diverge. The nodes are however closer to one another earlier in formation when the EE Draconis cluster had yet to settle into its position centrally located in its node. This may suggest that formation was originally more intimately linked between the two groups, with the division into nodes being an event that took place after star formation had already started. The second notable feature is the wide range in ages across both nodes, with age solutions spread broadly across the period that appears to have been active. These facts combined seem to present a picture of a continuous but internally divided star formation history in CFN.  

\subsection{Total Mass and Population} \label{sec:completeness}

Through our population of known members combined with our search for binaries and known binary and system demographics, we can estimate the number of stars and amount of stellar mass that is unaccounted for in our current survey of CFN members. We began with the sample of 549 candidate CFN members, which was drawn from a wider sample of 901 stars, of which 53$\pm$5\%, or 478$\pm$45, are expected to be genuine members as described by photometric selection priors. This result is however not necessarily uniform across the entire CMD, as the removal of stars in this sample is done by purely photometric means, meaning stars on the pre-main sequence are much more likely to be true members. We therefore divide the sample into two sets for the purposes of setting membership probability: one for sources that can be assigned as young based purely on their photometry, and another for stars with ambiguous photometry. The stars identified as young based on their photometry are assumed to be real members, while the remaining stars with ambiguous photometry are taken to have a certain probability of membership. With 163 stars with ambiguous photometry and 478 expected members, we conclude that 92 of the stars with ambiguous photometry must be members, or 56\%. We therefore apply a corrective factor of 0.56 to exclude the likely contribution from field contamination in this sample with ambiguous photometric youth. Since we lack coverage of objects less massive than 0.09 M$_\odot$, as they were not included in SPYGLASS-I, we also apply a corrective factor according to integrated sum of the \citet{Chabrier05} initial mass function below this mass cutoff. By integrating the IMF from 0.09 M$_\odot$ to the brown dwarf mass cutoff at $\sim 0.01 $M$_\odot$, we find that approximately 2\% of mass is locked in these low-mass stars, and therefore a corrective factor of 1.02 reintroduces their mass into the sample. 

We must now add in stellar companions. Our search for binaries has already found many such companions, especially those that are cleanly separated, however it will not be complete. Some stars with small separations, especially those with large flux ratios will not always trigger our RUWE binarity flag \citep{Belkurov20,Wood21}. We therefore exclude stars in our population that were marked by our binaries search as secondaries, which allows us to estimate the amount of mass in companions based on the population of primaries alone. To do this, we use the updated multiplicity rates as a function of mass generated in \citet{Sullivan21}, interpolating between the masses with multiplicity estimates based on the mass of each star. We then use the \citet{Sullivan21} power law indices for companion mass ratio to compute mean mass ratios for each power law, again interpolating between the masses provided to estimate the mean mass ratio for each star in our sample. By multiplying the mean mass ratio by the multiplicity rate and mass of the primary, we generate a mean missing mass for each primary. A similar approach can be used to estimate the number of missing secondaries, which can be done by summing the multiplicity rates across all stars in the sample. 

By combining our missing stars and stellar mass estimates with our corrections for contamination and mass in low-mass stars, we calculate that CFN contains a total of about $505\pm48$ stars spanning $359\pm34$ systems and containing a total mass of $242\pm23$ M$_\odot$. We also provide regional mass estimates, which are included in Table \ref{tab:sgprops}. The uncertainties provided are based on the 9\% fractional uncertainty in the $53\pm5$\% prior which conditions the key assumptions of our population estimates. However, there are other uncertainty sources that are much more difficult to quantify, most notably systematic model-based uncertainties in mass estimation that may affect our mass results \citep[e.g., see][]{Rizzuto15,Feiden16}, as well as uncertainties in our reintroduction of binary systems due to uncertainties in their demographics at young ages \citep[e.g.,][]{Kraus11,DeFurio19,DeFurio22}. These additional sources may raise the true uncertainties to over 10\%, however such an analysis is beyond the scope of this paper. 

There are also two ways in which populations in CFN may be systematically underestimated. First is the loss of members through the absence of a 5-parameter Gaia astrometric solution, an issue that typically manifests for relatively tight binaries (150-300 mas) with a contrast less than about one magnitude \citep{Wood21, Kraus23}. Using the \citet{Sullivan21} binary demographics combined with \citet{Raghavan10} information on the separation distribution, we however find that less than 2\% of the population lies in this range that risks being missed, making the expected effects minimal. 
The other potential cause of population underestimation is the likely presence of a limited population of CFN members beyond our selected boundaries. However, based on our observation that only 5\% of stars with $D<0.05$ have photometry consistent with youth, no more than 80 of those wider-population stars have a significant likelihood of youth. Furthermore, due to the heavy sample contamination farther from CFN's core, many of these possibly young stars are expected to be unresolved field binaries. While a 10-20\% increase to the population of CFN is possible when including extended populations, these stars are so widely distributed that they are unlikely to impact any gravitational binding assessments for subgroups of CFN, nor are they likely to reveal any new structural patterns.

\subsubsection{Is EE Dra a Cluster?} \label{sec:res-EEDra}

The expected masses of the populations within CFN can be used to compute the virial state of EE Dra, the one CFN subregion that shows evidence for a non-negligible internal gravitational potential in our traceback results. To investigate the binding state of the EE Draconis cluster, we begin by gathering basic properties of the group. First, we remove outliers. While EE Draconis has an overall extent of over 10 degrees on the sky, the vast majority of its mass is in a small core less than one degree across in the plane of the sky, which can be isolated by restricting selection to galactic longitude $99.5 <l< 101.2$ and latitude $b < 26$. We use the projected on-sky extent rather than the XYZ extent to avoid parallax uncertainties, which can produce lengthy extensions of features at EE Dra's distance $d = 193$ pc, especially relative to a sub-parsec extent on-sky. This remaining region has an expected mass of approximately 11.6 M$_{\odot}$, and a half-mass radius of 0.16 degrees, or 0.54 pc at the distance of EE Dra. 

Unbound systems satisfy the inequality $\sigma_{1D} > \sqrt{2} \sigma_{virial}$, with $\sigma_{1D}$ being computed as the square root of the mean variances in a multi-dimensional velocity dispersion \citep{Kuhn19}. The virial velocity, $\sigma_{virial}$, can be computed as follows \citep{Kuhn19}:

\begin{equation}
\sigma_{virial} = (\frac{G M}{r_{hm} \eta})^{-1/2}    
\end{equation}

where $r_{hm}$ is the half-mass radius, $M$ is the group mass, and $\eta$ is a function of the cluster's density profile, which is equal to 10 for a Plummer profile, but is often lower for the broader density profiles common in young clusters, varying by up to a factor of 3 \citep{PortegiesZwart10}.  Taking a relatively low value of $\eta = 5$, we compute $\sigma_{virial} = 0.13$ km s$^{-1}$, and using the more standard $\eta = 10$ we return 0.10 km s$^{-1}$. CFN-1, the most populous region in the association and densest outside EE Dra has $\sigma_{virial}$ values approximately 2.5 times smaller despite CFN-1's visibly wider velocity distribution, verifying that EE Dra is indeed an outlier within this association in terms of its internal gravitational potential. 

We must now calculate the value of $\sigma_{1D}$ for EE Dra for comparison to $\sigma_{virial}$. We compute this value based the same data set used for traceback, only including stars that pass our likely member cut and have no resolved binary companions, as binaries introduce additional velocity dispersion. The value itself is calculated based on the transverse velocity variances along the $l$ and $b$ axes, clipped at 2$\sigma$ to remove a handful of major outliers. Averaging the variances, we find a value of $\sigma_{1D} = 0.182$ km s$^{-1}$. We provide values of $\sigma_{1D}$ for the other subgroups in Table \ref{tab:sgprops}, including a solution for EE Dra without limiting the sample to its dense core. The combined uncertainties in the velocities that produce this value have a median of 0.041 km s$^{-1}$, so we subtract this value from $\sigma_{1D}$ in quadrature, producing an estimate for the intrinsic dispersion without the observational uncertainty of $\sigma_{1D} = 0.177$ km s$^{-1}$. This is between the values of $\sqrt{2} \sigma_{virial} = 0.19$ with $\eta = 5$ and $\sqrt{2} \sigma_{virial} = 0.14$ for  $\eta = 10$. Many of these values are approximate due to the small number statistics involved, especially $r_{hm}$, the transverse velocity dispersions, and the mass profile constant $\eta$, however these results are broadly consistent with a virialized state in which $\sigma_{1D} \approx \sqrt{2} \sigma_{virial}$. This represents a state of equilibrium, in which internal velocity vectors are expected to be dominated by random motions, rather than dispersal as is the case for associations. This matches with our conclusion from attempting traceback on the cluster, in which traceback does not produce a more compact configuration at an earlier time, indicating that dispersal is not the dominant source of motion in the association. We therefore conclude that the status of EE Dra as an essentially virialized cluster is supported by our knowledge of the group's stellar distribution. 

With its low mass and compact configuration, EE Dra represents a unique example of a cluster, with an incredibly compact configuration and low mass. The closest known analog to EE Dra is the $\eta$ Chamaeleontis cluster, which has a central core spanning less than 1 pc and a mass M$<20 $M$_\odot$ while also containing an extended halo of presumed ejected stars residing well beyond the dense core \citep{Mamajek99, Murphy12}, all facts that also apply to EE Dra. $\eta$ Cha also has an escape velocity of $v_{esc}\simeq0.5$ km s$^{-1}$ capable of binding its members, which have a SPYGLASS-I-based 1-D velocity dispersion of $\sigma_{1D} = 0.25$ km s$^{-1}$ \citep{Murphy12}. Even the namesake central stars are of similar luminosities, with both stars having Gaia magnitudes slightly dimmer than $G=0$. EE Dra therefore appears to represent a new case study in an emerging pattern of dense and virialized low mass populations in the solar neighborhood. With an age of over 25 Myr, this population is also much older than $\eta$ Cha ($\simeq$9 Myr; \citealt{Alecian07, Bell15, Kerr21}), greatly extending the period of time over which clusters like this are known to survive after formation.

\subsection{Disks}

\begin{figure}
\centering
\includegraphics[width=8cm]{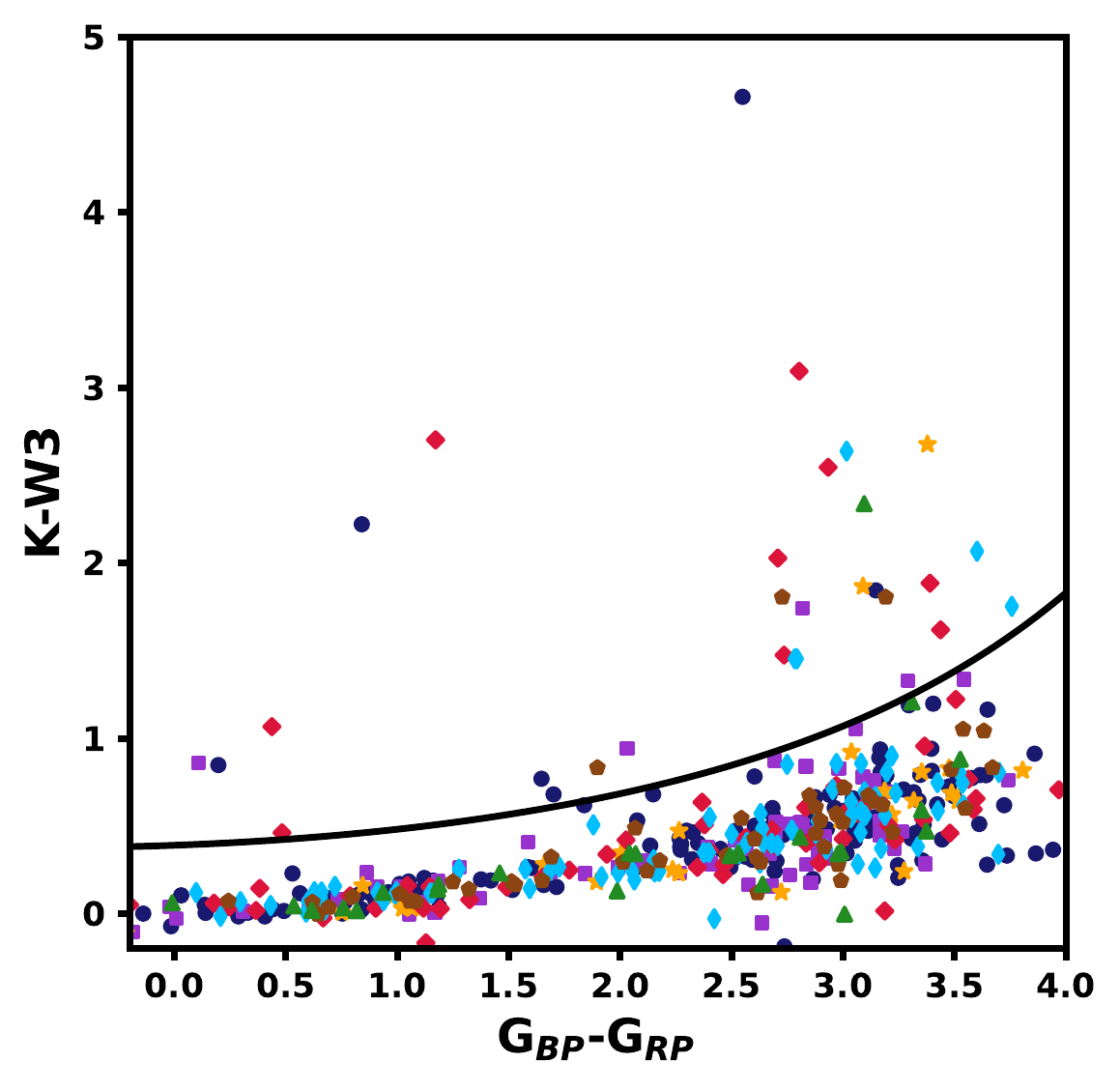}\hfill
\caption{K-W3 color excess for all stars in the high-confidence member sample, plotted against Gaia G$_{BP}$-G$_{RP}$ color. Stars above the black curve are identified as disk-bearing. Object icons use the same color and marker scheme first introduced in Figure \ref{fig:xyzclustering}. }
\label{fig:disks}
\end{figure}

Disks are another notable feature of young stellar populations, and while their abundances do not follow as predictable of a pattern with increasing age compared to spectroscopic youth indicators, they are nonetheless an important feature in tracking the advancement of star formation events. They are also important for studies of planet formation, both revealing a sample of environments of active formation, and through their presence or absence providing insight into the dispersal timescale of material capable of forming planets \citep[e.g.,][]{Ercolano17}. To identify disks, we performed a Vizier query of the ALLWISE catalog \citep{ALLWISE14} and retrieved the Ks and W3 magnitudes. We identified disks according to the dispersion of colors in K-W3 as a function of Gaia BP-RP color, clipped to avoid outliers and smoothed and extrapolated to get a steady cut across the full range of magnitudes explored. Stars more than 3-$\sigma$ above the median of this color-color sequence were taken to host disks. We only considered sources that had reported ALLWISE magnitude uncertainties, as numbers in ALLWISE that are reported without uncertainties are actually 2-$\sigma$ detection upper limits, rather than real magnitude measurements \citep{ALLWISE14}. We show our selection of disk candidates in Figure \ref{fig:disks}, alongside the cut that we use.

We performed this disk search on the full sample of 549 candidate members, producing 36 disk candidates out of 489 stars that have credible K-W3 photometry. All disk candidates are recorded in Table \ref{tab:members}. We made a further restriction to the set of 302 high-confidence members for additional analysis, as we found that restrictions to the astrometric and photometric quality flags in particular helped to remove many stars low on the pre-main sequence with more uncertain membership that may be part of a contaminated field. This new cut restricted the set to 17 disks across the 277 stars with credible WISE photometry. The effect of the level of restriction on disk fraction is however limited, as the candidate sample has a disk abundance of 6.1$^{+1.3}_{-1.1}$\%, and the high-confidence sample has a disk abundance of 4.7$^{+1.6}_{-1.3}$\%. The disk fraction for CFN-4 is a notable outlier at 15$^{+9}_{-6}$\%, which may be explained by its age, which at 16.1$\pm$1.4 Myr is the youngest in the association.

\section{Discussion} \label{sec:discussion}

\subsection{Stars Form in Multiple Distinct Nodes}

One particularly notable feature of our CFN traceback was the emergence of two nodes of star formation. These nodes currently overlap spatially, however our traceback of members along their galactic orbits (Figure \ref{fig:traceback}) reveals that these nodes have been separated by approximately 30 pc in the past, with no known interloping members. While formation in the two nodes progressed on largely the same timeline, the nodes had little interaction with one another for long periods of time while star formation was still active. 

One possible explanation for this coeval and dual-node structure is an origin in a large-scale filament. Early star formation in CFN resided along a somewhat non-linear path in spatial coordinates (as can be seen in Figure \ref{fig:traceback}), which could trace a somewhat curved filamentary structure. CFN's scale is quite consistent with known filaments, as the filamentary ``bone'' structures identified in \citet{Zucker15} have a length range roughly centered on this 30 pc length, a scale which is seen frequently throughout studies of dense filamentary structures \citep[e.g.,][]{Wang16, Zucker18}.  Many of those filaments have shapes that are sufficiently warped to fit the distribution of group centers seen around the time of CFN's first star formation event, and the masses that the filaments contain are also consistent with the production of a CFN-scale stellar population given a standard 1-10\% star formation efficiency \citep[e.g.,][]{Zucker15,Federrath13}. The 2-3 km s$^{-1}$ velocity differences between CFN nodes are also consistent with a filamentary origin, as velocity gradients on that level are common in observed filaments \citep{Hacar22}. The division into nodes could then be the result of cloud fragmentation, which has been shown to occur at various scales along filaments, or through accretion onto multiple collection hubs along the filament \citep[e.g.,][]{Hacar22, Beuther15}.

It is however worth noting that offsets in formation positions between subgroups have also been shown in simulations of spherical clouds, without the need to invoke filamentary formation. The STARFORGE simulations from  \citet{Grudic22} and \citet{Guszejnov22} represent the largest currently available star formation models capable of resolving the full spectrum of the IMF, and they include all of the major physical processes involved in star formation (i.e. magneto-hydrodynamics, radiative heating and cooling, proto-stellar jets, stellar winds, radiative and ionizing feedback, MS and pre-MS stellar evolution, supernovae explosions; \citealt{Grudic21}). An assessment of clustering within these simulations in \citet{Guszejnov22} frequently showed clusters which are separated by around 10 pc emerging from a 10 pc radius spherical cloud, demonstrating that this type of spatially dispersed substructure does not require any particular initial cloud shape. That said, the fact that the spatial separations seen between the CFN nodes are larger than seen in these simulations, exceeding 20 pc, and that there are no clear indications of any formation outside these nodes does hint that emergence from a single isotropic collapse is unlikely, and that the initial cloud was likely at least somewhat elongated. Given how common filamentary structures are in known gas clumps \citep[e.g.,][]{Andre10,Wang16,Zucker18}, the production of CFN in an elongated filament seems probable.

\subsection{The History is Consistent with Continuous Formation in Each Node}

\edit1{The $\beta$ Cep and EE Dra nodes show star formation spanning 7 and 9 Myr respectively}, with age gaps between subsequent generations of \edit1{$\sim$1-3 Myr} for all but one pair of populations. This represents a fairly even spread of ages across the period of active formation, and since stars within individual populations often have 3-5 Myr spreads against isochrones (see Figure \ref{fig:isoagefits}), this suggests that CFN's star formation history is continuous rather than bursty. A more complete assessment of whether CFN's age distribution implies continuous formation requires a detailed study of the age spreads within each population we identify. Accurate measurements of these age spreads require a detailed understanding of the uncertainties in photometry, parallax, reddening, and cross-contamination between populations to be meaningful, making this analysis beyond the scope of this publication. While it is possible that future studies reveal these spreads to be dominated by uncertainties and contamination and that star formation pauses are present, the consistently small age gaps between generations, which are generally consistent with the duration of small star formation events in current models \citep[e.g.,][]{Grudic20,Guszejnov22}, support the conclusion that star formation was continuous. 

A continuous star formation history may have somewhat different implications between the two nodes. The $\beta$ Cep node shows effectively co-spatial formation between subsequent generations, suggesting that what we see today may be less the result of distinct populations and more a single, continuous event with different generations differentiable only due to the different velocities produced according to their positions in the galactic potential at the time of decoupling from the molecular cloud. The EE Draconis node is much more loose, with clear 10-20 pc spatial separations between subgroups at the times of formation that make physical interactions between subsequent generations less likely. This may imply that the EE Draconis node did not form out of a single gas collection hub, instead forming from part of a parent filament that fragmented further into three different hubs, each hosting a small but separate star formation event.
Each hub would have collapsed at its own rate, starting with the EE Dra cluster, which is the densest of these groups in the present day. This sort of hierarchical fragmentation would look not unlike the spatially dispersed collection of gas clumps and subsequent clusters that emerged in the \citet{Guszejnov22} cluster analysis of STARFORGE simulations, as our EE Dra subgroups have a similar level of spatial separation while also hosting similar scales to what these simulations show. Future comparisons to simulations are necessary to further assess the implications of our results in CFN on current models of cluster formation and assembly. 

As a relatively low-mass environment, our reconstruction of star formation in CFN does not yet have clear parallels in the observational literature. Continuous star formation events have previously been shown, although they tend to come alongside a more consistent spatial trend, like the sequential star formation SPYGLASS-I showed in Sco-Cen. Our traceback of CFN does not indicate any spatial propagation of star formation, however there is still a spatial distribution in ages, albeit not nearly as intuitive as what was shown in Sco-Cen. A closer analog to star formation in CFN may be found in Taurus, where older generations appear present alongside multiple distinct filaments where star formation is ongoing \citep{Kraus17, Krolikowski21}. While the populations known in Taurus have much more substructure compared to CFN \citep[e.g., see][]{Krolikowski21}, part of that is likely caused by its younger age not providing the time necessary to blend populations. CFN could therefore provide insight into what a population like Taurus could look like 15-20 Myr in the future. 
The results in CFN nonetheless demonstrate that non-sequential star formation can produce star formation that appears sequential, motivating a need for studies like this one in a wide range of associations. 
Star formation pauses tend to emerge in higher-mass environments, which makes sense given the presence of O stars capable of quenching star formation via feedback processes \citep[e.g.,][]{Beccari17, Zhou22}.
With Cepheus Far North being a relatively dispersed low mass environment with fewer high-mass stars than any of these aforementioned examples, the apparent lack of star formation pauses is perhaps not surprising, however additional studies in comparable regions like Carina-Musca will be necessary to establish whether a history like what we see in CFN is typical for associations in this mass range.

\subsection{Reflections on SPYGLASS-I}

This paper provides the first of several planned detailed follow-up analyses to associations first covered in SPYGLASS-I. The characterization of associations in SPYGLASS-I was limited in nature, as it did not include photometrically older candidates in clustering, and it lacked the precise radial velocities and multi-method age solutions that we employ in this publication. This work therefore represents a significant improvement over SPYGLASS-I in terms of clustering sensitivity and age precision while also introducing traceback, which greatly improves our ability to assess connections between subregions during formation. With our new and much more detailed look at CFN, it is valuable to investigate how our results for CFN compare to those in SPYGLASS-I, and what implications those differences might have for claims made concerning other SPYGLASS-I associations. 

SPYGLASS-I's results in CFN depicted a much less complex view of the association, identifying only two subgroups. Our results here show much more substructure especially in the SPYGLASS-I-defined CFN-1 subgroup, which this work subdivides into six additional subgroups. While maximally sensitive association-level subclustering was not a core element of the analysis in SPYGLASS-I, the presence or absence of age differences was highlighted. In the case of CFN and a few other associations, such as Carina-Musca and Monoceros Southwest, SPYGLASS-I found a consistency in the subgroup ages spanning areas of order 100 pc, suggesting the presence of large-scale coeval star formation. The detailed results we present here in CFN however complicate that narrative by revealing that the CFN-1 subgroup as defined in SPYGLASS-I actually contains a distinct range of ages on either side of the bulk age solution computed in SPYGLASS-I. Rather than revealing coeval star formation over an entire large association, our results here show distinct subgroups and small-scale age variations which are averaged over in SPYGLASS-I. 

Our revised clustering and age results in CFN therefore further motivate follow-up studies in associations like Carina-Musca, where very consistent ages were measured across the two subregions identified, but with one of the subregions (CM-2) having visible substructure which might hide age variations. This structure may suggest a similarity between CM-2 and SPYGLASS-I's definition of CFN-1, in that an apparently contiguous and coeval population is identified, but multiple differently-aged subgroups exist within. Such an update would have a significant effect on our understanding of the region, as it would convert the SPYGLASS-I view of large-scale coeval formation into the continuous but gradual view of star formation that emerged in CFN. 
Improving clustering may have the opposite result in a region like Perseus, where SPYGLASS-I identified populations with wide age separations. Finer clustering may show a spread in age across subgroups within each currently-defined cluster, potentially capable of bridging the gap between the populations known. A study like what we provide for CFN would therefore be valuable in groups with properties similar to both Carina-Musca and Perseus, allowing us to investigate whether our observations of continuous star formation in CFN are reflected in similarly populated regions, or whether star formation bursts or large-scale rapid star formation can take place during the formation of these relatively low-mass associations.

\section{Conclusion} \label{sec:conclusion}

We have provided the first in-depth view of the structure, dynamics, and star formation history of Cepheus Far North. Using Gaia astrometry and photometry combined with additional spectroscopy, we compute ages and perform dynamical traceback on the association's members, thereby unraveling the star formation history of CFN and providing an unprecedented view of its structure. The key discoveries made through these investigations are as follows:

\begin{enumerate}
    \item We perform a detailed census of the CFN association, identifying 549 candidate members including 302 high-confidence members, a significant expansion over SPYGLASS-I (219 members). We also estimate the association's total population, which we find contains $\sim$505 stars across $\sim$359 systems, with a total stellar mass of $\sim$242 M$_{\odot}$. 

    \item Out of this stellar population, we identify seven spatially and dynamically coherent subgroups in CFN. These populations are spread across two separate nodes containing stars which were co-spatial throughout much of their star formation history, one that began with the formation of EE Dra, and the other with $\beta$ Cep. 

    \item Using a combination of dynamical, isochronal, asteroseismic, and lithium depletion ages, we find that star formation in CFN spans approximately 10 Myr, with the EE Dra node forming between \edit1{17} and 26 Myr ago, and the $\beta$ Cep node forming between 16 and \edit1{23} Myr ago. The small age gaps between populations is consistent with continuous star formation in each node, rather than a bursty history. 
    
    \item We identify a new open cluster around EE Draconis, which we name after that star. With a total mass of $\sim$16 M$_{\odot}$ and core spanning less than 1 pc, it is one of the smallest and least massive open clusters ever discovered, with properties consistent with being an older ($25.8\pm2.7$ Myr) analog to the $\eta$ Cha cluster.  

    \item We find that our more sensitive clustering analysis in CFN reveals significant age structure that did not emerge on larger scales. This result suggests that age variations within associations exist on smaller scales than previously thought, suggesting that previous assertions of large-scale coeval star formation may have averaged over the age differences across smaller-scale subgroups. 
\end{enumerate}

CFN is a small and sparsely populated association, and even in this supposedly simple environment we observe a wealth of substructure and a rich star formation history. This is a strong indication that many of the lesser-known  populations that have emerged in SPYGLASS-I have a significant story to tell, and as more associations like it are discovered, new attention from the community will be necessary to investigate the wealth of new populations that emerge, establish patterns, and reveal what these populations can tell us about star formation. 

\begin{acknowledgments}

RMPK thanks Gregory Feiden, who gave us access to the DSEP-Magnetic models. RMPK is funded by the Heising-Simons Foundation. RMPK acknowledges the use of computational  resources at  the  Texas  Advanced Computing Center (TACC) at the University of Texas at Austin, which was used for the more computationally intensive operations in this project, most notably our extraction of spectra and RV solutions. SJM is funded by the Australian Research Council through Future Fellowship FT210100485.

\end{acknowledgments}

\vspace{5mm}
\facilities{Gaia, Las Cumbres Observatory: NRES Spectrograph on 1m Telescopes at McDonald Observatory and Wise Observatory, McDonald Observatory: Tull Coud\'e Spectrograph at the 2.7 m Harlan J. Smith Telescope}


\software{astropy \citep{Astropy13},  
          {\tt saphires} \citep{Tofflemire19},
          echelle \citep{hey&ball2020}
          }

\appendix 

\section{Binaries} \label{app:bin}

\begin{deluxetable*}{ccccccc} 
\tablecolumns{7}
\tablewidth{0pt}
\tabletypesize{\scriptsize}
\tablecaption{A catalog of binaries in CFN. Objects identified as members of the same system are given the same system ID. }
\label{tab:binaries}
\tablehead{
\colhead{Gaia ID} &
\colhead{Sys ID} &
\colhead{$R$\tablenotemark{a}} &
\colhead{RA} &
\colhead{Dec} &
\colhead{m$_G$} &
\colhead{$\pi$} \\
\colhead{} &
\colhead{} &
\colhead{AU} &
\colhead{(deg)} &
\colhead{(deg)} &
\colhead{} &
\colhead{(mas)}
}
\startdata
 2280112203742060928 &          0 &      0 &  329.6419 &  75.0547 &  10.20 &      5.85 \\
 2280112208035806336 &          0 &   1338 &  329.6445 &  75.0568 &  16.21 &      5.89 \\
 2276738081729485696 &          1 &      0 &  316.9389 &  73.5925 &  14.86 &      6.01 \\
 2276738077434022016 &          1 &   2310 &  316.9525 &  73.5930 &  18.17 &      6.16 \\
 2277005533636485376 &          2 &      0 &  315.2675 &  74.2236 &  11.23 &      6.15 \\
 2277005537933326464 &          2 &    432 &  315.2689 &  74.2230 &  16.41 &      5.73 \\
 2278408411691360768 &          3 &      0 &  318.0162 &  76.3102 &   7.10 &      4.55 \\
 2278408308612143872 &          3 &   7018 &  317.9819 &  76.3150 &  15.97 &      4.84 \\
 2278408308612145408 &          3 &   2576 &  318.0059 &  76.3079 &  11.18 &      4.57 \\
 2281573729572210432 &          4 &      0 &  345.1857 &  77.4774 &  10.81 &      6.00 \\
 2281573733870090240 &          4 &    261 &  345.1860 &  77.4777 &  14.94 &      6.03 \\
 2281640494842167040 &          5 &      0 &  350.9689 &  77.0691 &   9.18 &      5.30 \\
 2281640494842166528 &          5 &   1286 &  350.9648 &  77.0708 &   9.32 &      5.28 \\
\enddata
\tablenotetext{a}{Separation at the distance of the primary relative to the primary. Primaries have a separation of zero.}
\vspace*{0.1in}
\end{deluxetable*}

Here we provide a complete catalog of binary and multiple systems identified in CFN, covering all objects within 10000 AU of each star in the plane of the sky at the star's distance, provided that they have parallaxes within 20\% and proper motions within 5 km s$^{-1}$. When other CFN members were included in the search radius around a star, the results of a search were only recorded if the star being searched was brighter than all other CFN members. This choice avoids duplicates in the list and focuses the search around the primary, which is typically more central in the system. All systems are assigned a system ID, which tracks the parent system of each component. Our resulting catalog contains 163 stars associated with a binary or multiple system, including 21 stars not in our candidate member catalog, likely left out due to their high internal velocity produced from interaction with the companion. 

\bibliography{sample631,sjm_bib}{}
\bibliographystyle{aasjournal}

\end{document}